\documentclass[11pt,a4paper]{article}
\pdfoutput=1 

\usepackage{jcappub} 

\usepackage[T1]{fontenc} 
\usepackage{dcolumn}
\usepackage{booktabs}
\usepackage{multirow}

\title{\boldmath Dark matter relic density\\
in scalar-tensor gravity revisited}

\author{Michael T. Meehan}
\author{and Ian B. Whittingham}
\affiliation{College of Science, Technology and Engineering, James Cook University,\\ Townsville 4811, Australia}

\emailAdd{Michael.Meehan@my.jcu.edu.au}
\emailAdd{Ian.Whittingham@jcu.edu.au}

\abstract{We revisit the calculation of dark matter relic abundances in scalar-tensor gravity using a generic form $A(\varphi_*) = e^{\beta\varphi_*^2/2}$ for the 
coupling between the scalar field $\varphi_*$ and the metric, for which detailed Big Bang Nucleosynthesis constraints are available. We find that BBN constraints restrict the modified expansion rate in these models to be almost degenerate with the standard expansion history at the time of dark matter decoupling. In this case the maximum level of enhancement of the dark matter relic density was found to be a factor of $\sim 3$, several orders of magnitude below that found in previous investigations.}

\begin{document}
\maketitle
\flushbottom

\section{Introduction}
\label{sec:intro}

Combined observations of large scale structure, galaxy dynamics and the cosmic microwave background (among others)~\cite{Hooper} now provide overwhelming evidence for the existence of non-baryonic dark matter with present density ($68\%$ C.L.)~\cite{Planck:2015xua}
\begin{equation}
\Omega_{DM} = 0.1188\pm 0.0010\,h^{-2},\label{eq:dm_abun}
\end{equation}
where $\Omega_{DM}$ is the dark matter density as a fraction of the total mass-energy budget and $h = 0.6774\pm 0.0046$ is defined by the present value of the Hubble constant $H_0 = 100\, h$ km/s/Mpc. 

In spite of this evidence, dark matter particles have never been observed directly and a convincing description of their particle nature remains elusive. The data favour cold (non-relativistic) dark matter with WIMPs (Weakly Interacting Massive Particles) the most popular theoretical candidate.\footnote{Within this class lies the widely studied neutralino, the lightest supersymmetric particle in supersymmetric extensions of the Standard Model in which $R$-parity is conserved.}

A popular framework for the origin of dark matter is provided by the thermal relic scenario in which the dark matter particles are produced through thermal scatterings of background particles in the cosmic bath (see e.g.~\cite{KandT}). At early times, when the temperature of the universe is high, frequent interactions keep the dark matter particles in equilibrium with the background. As the universe expands and cools the dark matter interaction rate drops below the expansion rate, the particles fall out of equilibrium, creation and annihilation processes cease and the number density redshifts with expansion. The surviving 'relic' particles constitute the dark matter density we observe today. This process is known as particle freeze-out and, for non-relativistic dark matter particles, typically occurs at $T_f \sim m_\chi/20$ (where $m_\chi$ is the dark matter particle mass).

Whilst in equilibrium, the dark matter number density, $n_\chi$, decays exponentially (i.e. $n_\chi \approx n_\chi^{\mathrm{eq}}\sim e^{-m_\chi/T}$ where $T$ is the temperature of the universe) so that the present dark matter abundance depends sensitively on the timing of freeze-out: the longer the species remains in thermal contact with the background bath, the lower its density at freeze-out. In the standard cosmological model of cold dark matter (CDM) with a non-zero cosmological constant ($\Lambda$), denoted the $\Lambda$CDM model, particle freeze-out occurs during the radiation dominated era when the expansion rate $H \sim T^2/M_{\mathrm{Pl}}$ (where $M_{\mathrm{Pl}} = 1.22\times 10^{19}$ GeV is the Planck mass). In this scenario, a dark matter species with a weak scale interaction cross section, $\sigma \sim G_{\mathrm{F}}^2\,m_\chi^2$, freezes out with an abundance that matches the presently observed value~\eqref{eq:dm_abun}. This is known as the 'WIMP miracle' and strongly motivates thermal WIMP dark matter models. 

Despite the observational success of $\Lambda$CDM, current datasets leave the physics of the universe prior to Big Bang Nucleosynthesis (BBN) 
($t \sim 200$ s) relatively unconstrained. If the universe experiences a non-standard expansion law at early times, and in particular during the era of dark matter decoupling, particle freeze-out may be accelerated (or delayed) and the relic abundance enhanced (or suppressed)~\cite{DBarrow1982501,PhysRevD.81.123522,Pallis:2009ed,Salati2003121,Arbey200846,Gelmini:2013awa,Iminniyaz:2013cla,Meehan:2014zsa,Meehan:2014b} (see also~\cite{Gelmini:2009yh}). 

In this article we investigate the relic abundance of dark matter in scalar-tensor theories of gravity~\cite{Jordan1949,Fierz:1956zz,Brans1961,Bergmann1968,1970ApJ...161.1059N,PhysRevD.1.3209,PhysRevD.61.023507} whose cosmological evolution deviates from the standard expansion history in the pre-BBN era. These theories are (in part) motivated by higher dimensional unification models, {\`a} la Kaluza-Klein~\cite{kaluza1921unitatsproblem,1926ZPhy...37..895K}, where additional scalar fields arise through the compactification of the higher dimensions and couple to the metric with gravitational strength. As such, the gravitational interaction is mediated by both the metric and scalar fields so that scalar-tensor gravity models represent a departure from standard General Relativity (GR).

Importantly, the new long range interaction introduced by the coupling, $A(\varphi_*)$, between the scalar field, $\varphi_*$, and the metric, $g_{\mu\nu}$,\footnote{Matter fields $\Psi_{\mathrm{mat}}$ couple directly to the so-called \textit{Jordan frame} metric $g_{\mu\nu} = A^2(\varphi_*)g_{\mu\nu}^*$, where an asterisk is used to denote \textit{Einstein frame} quantities, the frame in which the gravitational field equations take their standard form (see later).}  is subject to stringent experimental bounds from fifth force searches and solar system tests of gravity~\cite{Will:2014xja}. To evade detection in high density environments non-linear effects act to shield the scalar field through one of several screening mechanisms such as the Vainshtein or chameleon mechanisms~\cite{Khoury2010}.\footnote{For a recent paper on laboratory searches for the chameleon (scalar) field see~\cite{2015arXiv150203888H}.} In the chameleon case for example, the mass of the scalar field is background dependent so that in regions of high density (e.g. our solar system) the field mass is large and the interaction range is suppressed. Conversely, in low density backgrounds (i.e. on cosmological scales) the field can be extremely light, allowing the scalar interaction to drive the accelerated expansion of the universe.

In addition to the screening mechanisms that help shield the scalar field from astrophysical observations, many scalar-tensor gravity models exhibit an inherent attraction mechanism towards General Relativity~\cite{PhysRevLett.70.2217,PhysRevD.48.3436}. Throughout its cosmological evolution, the scalar field is driven towards a state where the coupling $A(\varphi_*)$ remains constant so that the scalar-tensor theory is indistinguishable from GR (see section~\ref{sec:scalar_dynamics}). This allows scalar-tensor gravity models to deviate from the standard cosmological scenario at early times whilst relaxing towards General Relativity prior to the onset of Big Bang Nucleosynthesis. Hence, these models could potentially disturb the timing of dark matter decoupling and, in turn, the predicted dark matter density.


The effects of the modified early time expansion rate in the scalar-tensor scenario on the dark matter relic abundance were first studied by Catena~\textit{et al} in~\cite{PhysRevD.70.063519}. In their paper the authors determined the evolution of the scalar field for the coupling $A(\varphi_*) = 1 + Be^{-\beta\varphi_*}$ and found that, for $T$ greater than some transition temperature $T_{\varphi_*}$, the early time expansion rate was enhanced by a factor of
\begin{equation}
\xi(T) = \frac{H_{\mathrm{ST}}}{H_{\mathrm{GR}}} \simeq 2.19\times 10^{14}\left(\frac{T_0}{T}\right)^{0.82}\qquad (\mbox{where}\;T_0 = 2.35\times 10^{-13}\,\mbox{GeV})\label{eq:xi_catena}
\end{equation}
before rapidly dropping to 1 at $T = T_{\varphi_*}$ (as a result of the attraction mechanism just mentioned). Interestingly, the authors discovered that the rapid relaxation of the scalar-tensor expansion rate, $H_{\mathrm{ST}}$, towards the standard expansion rate, $H_{\mathrm{GR}}$, at $T_{\varphi_*}$ led to a phase of \textit{reannihilation}: after the initial particle decoupling, the dark matter species experienced a subsidiary period of annihilation as the expansion rate of the universe dropped below the interaction rate. Despite this secondary annihilation phase, they showed that the relic abundance of dark matter in scalar-tensor gravity models can still be enhanced by up to three orders of magnitude (depending on the WIMP mass $m_\chi$). 

The approximate form~\eqref{eq:xi_catena} of the ratio $H_{\mathrm{ST}}/H_{\mathrm{GR}}$ has since been adopted in several subsequent investigations, including those by Gelmini~\textit{et al}~\cite{Gelmini:2013awa}, Rehagen and Gelmini~\cite{1475-7516-2014-06-044} and, very recently, Wang~\textit{et al}~\cite{Wang:2015gua} who studied the relic abundance of asymmetric dark matter species~\cite{Wang:2015gua,Gelmini:2013awa} and sterile neutrinos~\cite{1475-7516-2014-06-044}, and naturally obtained similarly large enhancement factors for the dark matter relic abundance. Subsequently  
Catena~\textit{et al}~\cite{PhysRevD.81.123522} considered the coupling $A(\varphi_*) = 1 + b\varphi_*^2$ as part of a more general study of dark matter relic abundances in non-standard cosmological scenarios and once again found that the relic abundance in scalar-tensor cosmology is enhanced by several orders of magnitude.

We should emphasize, however, that although the different couplings used by the Catena group allow for significant enhancements of the early time expansion rate, detailed studies of the BBN implications of scalar-tensor theories with these couplings are lacking and only the simple constraint 
\begin{equation}
\frac{A(\varphi_*)|_{T = 1\,\mathrm{MeV}}}{A(\varphi_*)|_{T = T_0}} < 1.08
\end{equation} 
was imposed.\footnote{This constraint on the coupling at the time of Big Bang Nucleosynthesis ($T = 1$ MeV) was derived from a constraint on the number of additional neutrino species $\Delta N < 1$.} However, detailed BBN studies are available (see for instance~\cite{PhysRevD.59.123502,0004-637X-658-1-1}) for the more popular choice of the scalar coupling 
\begin{equation}
A(\varphi_*) = e^{\frac{1}{2}\beta\varphi_*^2}.\label{eq:quad_coup}
\end{equation}
Thus it is important to determine if the relic abundance of dark matter can be similarly enhanced for these more widely investigated models. 

The quadratic coupling~\eqref{eq:quad_coup} is the simplest generalization of the original Jordan---Fierz---Brans---Dicke model (see later). 
Besides being the prototypical coupling considered in numerous investigations of scalar-tensor gravity theories, including studies of slowly rotating anisotropic neutron stars~\cite{Silva:2014fca} and pulsars in binary systems~\cite{PhysRevD.54.1474,PhysRevD.58.042001,freire2012relativistic}, the quadratic coupling was recently the subject of the rigorous investigation by Coc~\textit{et al}~\cite{PhysRevD.73.083525} in which BBN calculations were used to place stringent constraints on the various coupling parameters. These constraints are up to several orders of magnitude stronger than the solar system bounds (e.g.~\cite{Cassini}) and, as we will show in the following, severely limit the possible deviations from the standard cosmological expansion history during the era of dark matter decoupling.

Finally, we mention the 2008 paper by Catena~\textit{et al}~\cite{1126-6708-2008-10-003} where the authors introduced a new \textit{hidden} matter sector that experienced a different coupling than the ordinary \textit{visible} sector, thus allowing for slower pre-BBN expansion rates. This \textit{non-universal} scalar-tensor theory was found to predict dark matter relic abundances that were reduced by up to a factor of $\sim 0.05$. 

In this article we revisit the calculation of dark matter relic abundances in scalar-tensor gravity from first principles. We assume that the coupling between the scalar field and matter (including dark matter) is universal; the case of non-universal scalar-tensor theories, in which the coupling between the scalar field and the visible and dark matter sectors is distinct, will be briefly considered towards the end of the paper. We explicitly determine the modified expansion rate by solving the equation of motion for the scalar field directly and use this result to numerically solve the Boltzmann rate equation governing the evolution of the dark matter number density. We find, for the coupling $A(\varphi_*) = e^{\frac{1}{2}\beta\varphi_*^2}$, that the efficiency of the attraction mechanism towards GR combined with the strict BBN bounds on the input parameters only allow for modest enhancements of the dark matter relic abundance, particularly compared to those reported in Catena~\textit{et al}~\cite{PhysRevD.70.063519,PhysRevD.81.123522}. 

To begin, in section~\ref{sec:conformal_frames} we discuss the formulation of scalar-tensor theories in different conformal reference frames and comment on their physical interpretation. Then, after deriving the equations that govern the cosmological evolution in section~\ref{sec:st_cosmological}, we explore in detail the dynamics of the coupled scalar field and the attraction mechanism towards General Relativity (section~\ref{sec:scalar_dynamics}). We then investigate which regions of parameter space lead to the largest deviations from the standard expansion history (section~\ref{sec:st_modhub}) and, most importantly, which regions satisfy the bounds from Big Bang Nucleosynthesis and other astrophysical and dynamical constraints (section~\ref{sec:st_constraints}). In sections~\ref{sec:st_sdm} and~\ref{sec:st_adm} we calculate the dark matter relic abundance for symmetric and asymmetric dark matter models respectively, and, for the first time, determine the annihilation cross section required to produce the observed dark matter density. Then, in section~\ref{sec:st_nonuniversal}, we consider non-universal scalar-tensor theories and discuss the issues associated with determining the dark matter relic abundance when the scalar interaction with the standard matter (or visible) particle sector and dark matter particle sector is distinct. Finally in section~\ref{sec:st_summary} we summarize our results and comment on the potential for relic abundance calculations to discriminate between the scalar-tensor and standard cosmological scenarios.


\section{Scalar-tensor gravity: Jordan and Einstein frames}
\label{sec:conformal_frames}

Scalar-tensor gravity theories are often formulated in one of two conformal frames of reference, namely, the \textit{Jordan} and \textit{Einstein} frames.\footnote{Conformal reference frames are those connected by a conformal transformation, i.e. a local rescaling of the metric:
\begin{equation}
g_{\mu\nu} = \Omega^2(x) g_{\mu\nu}^*.
\end{equation}
Here, $\Omega(x)$ is an arbitrary function of the spacetime coordinates $x^\mu$, and we use the notation $\Omega^2$ to preserve the sign of the line element $ds^2 = \Omega^2 ds_*^2$.} The advantage of using these different conformal frames is that the scalar coupling enters through either the gravitational sector (Jordan frame) or the matter sector (Einstein frame), leaving the other sector unaffected. For example, the general action integral for a scalar-tensor theory with a non-minimally coupled scalar field $\varphi$, formulated in the Jordan conformal frame, is given by
\begin{equation}
S_{\mathrm{tot}} = \frac{1}{16\pi G_*}\int{d^4x\,\sqrt{-g}\left[F(\varphi)R - g^{\mu\nu}Z(\varphi)\partial_\mu\varphi\partial_\nu\varphi - 2U(\varphi)\right]} + S_{\mathrm{mat}}[g_{\mu\nu};\,\Psi_{\mathrm{mat}}],\label{eq:jord_act}
\end{equation}
where $F(\varphi)$, $Z(\varphi)$ and $U(\varphi)$ are arbitrary functions of the field, $G_*$ is the bare gravitational constant (i.e. the gravitational constant in the absence of the scalar interaction), and $R = g^{\mu\nu}R_{\mu\nu}$ is the Ricci scalar which has been constructed from the Jordan frame metric $g_{\mu\nu}$. 
The original Brans-Dicke model corresponds to $F(\varphi) = \varphi$, $Z(\varphi) = \omega/\varphi$ (where $\omega$ is a constant) and $U(\varphi) = 0$.

In this frame matter fields, $\Psi_{\mathrm{mat}}$, are coupled directly to the metric, $g_{\mu\nu}$, so that the Weak Equivalence Principle is preserved by construction. This means that observables such as mass, length and time take their standard interpretation in the Jordan frame, making it the most convenient for particle physics considerations. However, since in this frame the scalar field couples to the gravitational sector, gravitational couplings pick up an additional $\varphi$ dependence and the field equations take the cumbersome form:
\begin{equation}
F(\varphi)G_{\mu\nu} - \left(\nabla_\mu\nabla_\nu - g_{\mu\nu}\Box\right)F(\varphi) = 8\pi G_*\left(T_{\mu\nu} + T^\varphi_{\mu\nu}\right),\label{eq:jordan_field}
\end{equation}
where the covariant derivative $\nabla_\mu$, $\Box \equiv \nabla_\mu\nabla^\mu $ and Einstein's tensor, $G_{\mu\nu} = R_{\mu\nu} - \frac{1}{2}g_{\mu\nu}R$, are defined using the Jordan frame metric $g_{\mu\nu}$, and the matter and scalar field energy-momentum tensors are given respectively by
\begin{equation}
T_{\mu\nu} = -\frac{2}{\sqrt{-g}}\frac{\delta S_{\mathrm{mat}}}{\delta g^{\mu\nu}}
\end{equation}
and
\begin{equation}
8\pi G_* T_{\mu\nu}^\varphi = Z(\varphi)\partial_\mu\varphi\partial_\nu\varphi - g_{\mu\nu}\left[\frac{1}{2}g^{\alpha\beta}Z(\varphi)\partial_\alpha\varphi\partial_\beta\varphi + U(\varphi)\right].
\end{equation}

To simplify the gravitational field equations we could formulate the scalar-tensor theory in the 'Einstein' frame, which is related to the Jordan frame by the conformal transformation
\begin{equation}
g_{\mu\nu}^* = F(\varphi)g_{\mu\nu}.\label{eq:conf_tran}
\end{equation}
Here (and in the following) we use an asterisk to distinguish Einstein frame quantities from their Jordan frame counterparts. Applying the transformation~\eqref{eq:conf_tran}, the action~\eqref{eq:jord_act} becomes
\begin{equation}
S_{\mathrm{tot}} = \frac{1}{16\pi G_*}\int{d^4x\,\sqrt{-g_*}\left[R_* - 2g_*^{\mu\nu}\partial_\mu\varphi_*\partial_\nu\varphi_* - 4V(\varphi_*)\right]} + S_{\mathrm{mat}}[A^2(\varphi_*)g_{\mu\nu}^*;\,\Psi_{\mathrm{mat}}]\label{eq:ein_act}
\end{equation}
where we have made the following redefinitions
\begin{align}
\left(\frac{d\varphi_*}{d\varphi}\right)^2 &= \frac{3}{4}\left[\frac{d\ln F(\varphi)}{d\varphi}\right]^2 + \frac{Z(\varphi)}{2F(\varphi)},\nonumber\\
A(\varphi_*)&= F^{-1/2}(\varphi),\nonumber\\
2V(\varphi_*) &= \frac{U(\varphi)}{F^{2}(\varphi)}.\label{eq:field_redef}
\end{align}

The gravitational sector is now of the same form as a minimally coupled quintessence model. Moreover, the field equations take the simplified form
\begin{equation}
G_{\mu\nu}^* = 8\pi G_*\left(T_{\mu\nu}^* + T^{\varphi_*}_{\mu\nu}\right)\label{eq:ein_field}
\end{equation}
where the matter energy momentum tensor is now given by
\begin{equation}
T_{\mu\nu}^* = -\frac{2}{\sqrt{-g_*}}\frac{\delta S_{\mathrm{mat}}}{\delta g^{\mu\nu}_*},\label{eq:einem_mat}
\end{equation}
and the scalar field energy momentum tensor is
\begin{equation}
4\pi G_*T^{\varphi_*}_{\mu\nu} = \partial_\mu\varphi_*\partial_\nu\varphi_* - g_{\mu\nu}^*\left[\frac{1}{2}g^{\alpha\beta}_*\partial_\alpha\varphi_*\partial_\beta\varphi_* + V(\varphi_*)\right].\label{eq:einem_phi}
\end{equation}
However, in the Einstein frame, the scalar coupling enters through the matter action $S_{\mathrm{mat}} \equiv S_{\mathrm{mat}}[A^2(\varphi_*)g_{\mu\nu};\,\Psi_{\mathrm{mat}}]$ so that the matter fields $\Psi_{\mathrm{mat}}$ couple to the $\varphi_*$-dependent metric $A^2(\varphi_*)g_{\mu\nu}^*$. This indicates that particle physics quantities (e.g. mass, length, time, energies, cross sections) measured in this frame are spacetime dependent. 

We can characterize the departure of scalar-tensor theories from General Relativity by introducing the parameter
\begin{equation}
\alpha(\varphi_*) = \frac{d\ln A(\varphi_*)}{d\varphi_*}.\label{eq:alpha}
\end{equation}
Large values of $\alpha(\varphi_*)$ correspond to large variations in the coupling $A(\varphi_*)$ whereas in the limit $\alpha(\varphi_*)\rightarrow 0$, corresponding to $A(\varphi_*) = $ const., the Einstein and Jordan frames coincide and the scalar-tensor theory reduces to standard General Relativity.


In the following our strategy will be to determine the cosmological evolution in the Einstein frame, where the cosmological equations take their simplest form, and then transform our results over to the Jordan frame, which is where we solve the Boltzmann equation for the dark matter number density.


\section{Cosmological equations}
\label{sec:st_cosmological}

If we introduce the Einstein frame line element
\begin{equation}
ds_*^2 = g_{\mu\nu}^*dx_*^\mu dx_*^\nu = -dt_*^2 + a_*^2(t)\gamma_{ij}^*dx_*^i dx_*^j,
\end{equation}
and assume that the matter fields are a perfect fluid described by the usual energy-momentum tensor
\begin{equation}
T_{\mu\nu}^* = (\rho_* + p_*)u_\mu^* u_\nu^* + p_* g_{\mu\nu}^*,
\end{equation}
where $\rho_*$ and $p_*$ are the Einstein frame energy density and pressure of the fluid respectively, then the field equations~\eqref{eq:ein_field} give
\begin{align}
3H_*^2 &= 8\pi G_*\rho_* + \dot{\varphi}_*^2 + 2V(\varphi_*),\label{eq:ein_hub}\\
3\,\frac{\ddot{a}_*}{a_*} &= - 4\pi G_*\left(\rho_* + p_*\right) - 2\dot{\varphi}_*^2 + 2V(\varphi_*).\label{eq:ein_hub2}
\end{align}
Here an overdot denotes differentiation with respect to $t_*$ and we have introduced the Einstein frame Hubble factor $H_* \equiv d\ln a_*/dt_*$. Since the Friedmann equations~\eqref{eq:ein_hub} and~\eqref{eq:ein_hub2} depend on the scalar field, we also need the equation of motion for $\varphi_*$ to close the system and determine the evolution of the scale factor $a_*(t_*)$. 
Variation of the action~\eqref{eq:ein_act} with respect to $\varphi_*$ gives\footnote{Also needed is the relation~
\begin{equation}
\frac{\delta S_{\mathrm{mat}}[A^2(\varphi_*)g_{\mu\nu}^*,\Psi]}{\delta\varphi_*} = -\sqrt{-g_*}\alpha(\varphi_*)T_{\mu}^{*\,\mu}.\nonumber\\
\end{equation}}
\begin{equation}
\ddot{\varphi}_* + 3H_*\dot{\varphi}_* + \frac{\partial V}{\partial\varphi_*} = -4\pi G_*\alpha(\varphi_*)\left(\rho_* - 3p_*\right).\label{eq:ein_phimotion}
\end{equation}
Additionally, the Einstein frame energy density and pressure no longer satisfy the standard continuity equation. Instead there is an additional source term due to the scalar field interaction:
\begin{equation}
\frac{d\rho_{*i}}{dt_*} + 3H_*\left(\rho_{*i} + p_{*i}\right) = \alpha(\varphi_*)\left(\rho_{*i} - 3p_{*i}\right),\label{eq:ein_conservation}
\end{equation}
where $i$ labels the various fluid components (e.g. radiation, baryonic matter, dark matter). Instead of solving this equation directly we can transform to the Jordan frame in which
\begin{equation}
\frac{d\rho_{i}}{dt} + 3H\left(\rho_{i} + p_{i}\right) = 0,\label{eq:jord_conservation}
\end{equation}
where the Jordan frame expansion rate, $H = d\ln a/dt$ is related to the Einstein frame expansion rate via
\begin{align}
H &= A^{-1}(\varphi_*)\left[H_* + \alpha(\varphi_*)\dot{\varphi}_*\right].\label{eq:hjord}
\end{align} 
Note that~\eqref{eq:jord_conservation} can be derived from~\eqref{eq:ein_conservation} by transforming to the Jordan frame coordinates
\begin{equation}
a = A(\varphi_*)a_*\,,\qquad dt = A(\varphi_*)dt_*,\label{eq:ein2jord}
\end{equation}
and substituting in the relationships for the Einstein frame energy density, $\rho_*$, and pressure, $p_*$,
\begin{equation}
\rho_* = A^4(\varphi_*)\rho\,, \qquad p_* = A^4(\varphi_*) p.\label{eq:einjordrho}
\end{equation}

Assuming the pressure and energy density of the $i$-th fluid component are related by $p_i = w_i\rho_i$, where $w_i$ is the equation of state parameter, we get
\begin{equation}
\rho_i(t) = \rho_i(t_0)\exp\left[-3\int_{t_0}^t H(1 + w_i)dt\right].
\end{equation}
Further, if we assume that $w_i$ is constant,
\begin{equation}
\rho_i = \rho_i^0\left(\frac{a}{a_0}\right)^{-3(1 + w_i)}
\end{equation}
where a (sub)superscript '0' denotes a quantity evaluated at the present epoch, i.e. $\rho_i^0 = \rho_i(t_0)$. Then, using~\eqref{eq:einjordrho}, we finally have
\begin{equation}
\rho_{*i} = \rho_{*i}^0\left[\frac{A(\varphi_*)}{A(\varphi_{*0})}\right]^{4 - 3(1 + w_i)}\left(\frac{a_*}{a_{*0}}\right)^{-3(1 + w_i)}
\end{equation}
where $\rho_{*i}^0 = A^4(\varphi_{*0})\rho_{i}^0$. Notice from the relationship~\eqref{eq:einjordrho} between the Einstein and Jordan frame energy densities and pressures that the equation of state parameter is a frame invariant quantity:
\begin{equation}
w_i = \frac{p_i}{\rho_i} = \frac{p_{*i}}{\rho_{*i}}.
\end{equation}


\section{Scalar field dynamics}
\label{sec:scalar_dynamics}

To determine the expansion rate of the universe in scalar-tensor gravity models we must solve the coupled system of equations~\eqref{eq:ein_hub}-\eqref{eq:ein_phimotion}. Fortunately, as pointed out by~\cite{PhysRevD.48.3436,PhysRevLett.70.2217}, the equation of motion for the scalar field can be decoupled from the system by transforming the evolution parameter from the Einstein frame time $t_*$ to $N\equiv \ln (a_*/a_{*0})$. In this case variables of the type $\dot{Q}$ transform as $\dot{Q} = Q'H_*$ (where a $'$ denotes differentiation with respect to $N$), so that~\eqref{eq:ein_hub} becomes
\begin{equation}
(3 - \varphi_*'^2)H_*^2 = 8\pi G_* \rho_* + 2V(\varphi_*).\label{eq:ein_hub_p}
\end{equation}
Since the potential term only affects the cosmological evolution at late times and does not play a significant role during the early universe period in which we are interested, i.e. $V(\varphi_*) \ll H_*^2 \sim \rho_{*}$, we set $V(\varphi_*) = 0$.\footnote{The choice $V(\varphi_{*})=0$ does not give a viable late time cosmology in that it cannot account for the recent accelerated expansion of our universe. Coc \textit{et al}~\cite{PhysRevD.73.083525} have studied the effects of including a cosmological constant (in either the Jordan or Einstein frames) tuned to the observed value of the cosmological constant density today and find that only the late time dynamics of the scalar field is affected. Moreover, their analysis shows that the constraints on the parameters $(\beta,|\alpha_0|)$ (see section~\ref{sec:st_constraints}) are only moderately affected by the inclusion of a non-zero potential term, and that adding a cosmological constant in the Jordan frame actually constricts the allowed region in the $(\beta,|\alpha_0|)$ plane. Alternatively, Mota and Winther~\cite{Mota} consider general chameleon models (of which the quadratic coupling considered here is a special case) and study the conditions under which the chameleon potential $V(\varphi_{*})$ possesses an attractor solution where the chameleon follows the minimum of the effective potential $V(\varphi_{*}) + \rho_{\mathrm{m}}A(\varphi_{*})$. The authors found that the field will typically settle at this minimum before the onset of BBN provided $m_{\varphi_{*}} \gg H$. However, a later study by Erickcek~\textit{et al}~\cite{Erickcek2014} concluded large changes to $m_{\varphi_{*}}$ occur near the minimum, casting doubt on chameleon models in general.} Then, substituting~\eqref{eq:ein_hub_p} into~\eqref{eq:ein_phimotion} we get 
\begin{equation}
\frac{2}{3 - \varphi_*'^2}\varphi_*'' + (1 - w)\varphi_*' = -\alpha(\varphi_*)(1 - 3w)\label{eq:phimotion_universal}
\end{equation}
where $w$ is the equation of state parameter of the total cosmic fluid:
\begin{equation}
w = \frac{p_{*\mathrm{tot}}}{\rho_{*\mathrm{tot}}} = \frac{p_{\mathrm{tot}}}{\rho_{\mathrm{tot}}}.
\end{equation}
Following the analogy given in~\cite{PhysRevD.48.3436}, the field $\varphi_*$ can be thought of as a particle-like dynamical variable with a velocity dependent mass, $m({\varphi_*}) = 2/(3 - \varphi_*'^2)$. In this instance the particle ($\varphi_*$) experiences simple damping and rolls down the potential $\propto \ln{A(\varphi_*)}$ towards the minimum, provided such a point exists. Hence, the late-time evolution of the field is reasonably straightforward given that the equation of state parameter during the matter dominated epoch is $w \approx 0 $ so that the forcing term on the right hand side of~\eqref{eq:phimotion_universal} is simply given by $-\alpha(\varphi_*)$. Therefore, if the function $\alpha(\varphi_*)$ possesses a zero with a positive slope, the field will be dynamically attracted towards the point $\alpha = 0$, which is precisely the GR limit. For the choice of coupling $A(\varphi_*) = e^{\frac{1}{2}\beta\varphi_*^2}$ considered here, this condition implies that $\beta > 0$.


At much earlier times, deep the in radiation era ($w \approx 1/3$), the forcing term in~\eqref{eq:phimotion_universal} is mostly ineffective and any initial velocity possessed by the field is rapidly damped away. This allows us to take as initial conditions
\begin{equation}
\varphi_{*{\mathrm{in}}} = \mbox{const}.\quad \mbox{and} \quad \varphi_{*{\mathrm{in}}}' = 0,
\end{equation}
where $\varphi_{*{\mathrm{in}}}$ and $\varphi_{*{\mathrm{in}}}'$ are the values of the field and its $N$-derivative at some initial point $N_{\mathrm{in}}$. 

Although $1 - 3w$ is $\approx 0$ throughout most of the early universe, there is an important effect that arises when each of the particle species in the cosmic fluid becomes non-relativistic~\cite{PhysRevD.48.3436}. As the temperature of the universe drops below the rest mass of each of the particle types they provide a non-zero contribution to the quantity $1 - 3w$. This activates the forcing term in~\eqref{eq:phimotion_universal} and displaces, or 'kicks' the field along the potential $\propto \ln A(\varphi_*)$. In this way the attraction mechanism towards GR is initiated during the very early moments of the universe ($T\sim m_t \sim \mathcal{O}(10^2)$ GeV, where $m_t$ is the mass of the top quark), prior to the onset of Big Bang Nucleosynthesis.

To explore this effect in more detail we first write the quantity $1 - 3w$ as
\begin{align}
1 - 3w &= \frac{\rho_{\mathrm{tot}} - 3p_{\mathrm{tot}}}{\rho_{\mathrm{tot}}}\nonumber\\
&= \frac{1}{\rho_{\mathrm{tot}}}\left\{\sum_A\left[\rho_A - 3p_A\right] + \rho_{\mathrm{m}}\right\}
\end{align}
where we have separated out the contribution from the relativistic particle species $A$ and the non-relativistic particle species $\mathrm{m}$ (for which $p_{\mathrm{m}} = 0$). Assuming that the total energy density is dominated by relativistic matter and radiation during the crossing of each mass threshold, then $\rho_{\mathrm{tot}} \simeq \rho_{\mathrm{r}} = \pi^2 g_{*\rho}(T)T^4/30$ where $g_{*\rho}(T)$ is the total number of relativistic degrees of freedom. The \textit{kick function}
\begin{equation}
\Sigma(T) \equiv \sum_A{\frac{\rho_A - 3p_A}{\rho_{\mathrm{tot}}}}
\end{equation}
takes the form
\begin{equation}
\Sigma(T) = \sum_A{\frac{15}{\pi^4}\frac{g_A}{g_{*\rho}(T)}z_A^2\int_{z_A}^\infty{dx\,\frac{\sqrt{x^2 - z_A^2}}{e^x \pm 1}}},\label{eq:sigma}
\end{equation}
where $z_A = m_A/T$ and $+(-)$ corresponds to fermions(bosons), since the energy density $\rho_A$ and pressure $p_A$ of each of the particles of type $A$ are given by (see e.g.~\cite{KandT})
\begin{align}
\rho_{A}(T) &=\frac{g_A}{2\pi^2}\int_{m_A}^\infty{\frac{\left(E^2 - m_A^2\right)^{1/2}}{\exp{\left(E/T\right)} \pm 1}E^2\,dE},\\
p_{A}(T) &=\frac{g_A}{6\pi^2}\int_{m_A}^\infty{\frac{\left(E^2 - m_A^2\right)^{3/2}}{\exp{\left(E/T\right)} \pm 1}\,dE}.
\end{align}
Here $g_{A}$ is the number of internal (spin) degrees of freedom.
It is important to note that the variable $T$ in the expression for $\Sigma(T)$ is the Jordan frame temperature, which we can relate to the parameter $N$ through
\begin{equation}
T[\varphi_*,N] = T_0\frac{A(\varphi_{*0})}{A(\varphi_*)}\left[\frac{g_{*s}(T_0)}{g_{*s}(T)}\right]^{1/3}e^{-N}.\label{eq:TtoN}
\end{equation}
Here $T_0 = 2.35\times 10^{-13}$ GeV is the current temperature of the universe and $\varphi_{*0} = \varphi_*(T_0)$. 

\begin{figure}[htbp]
\centering
\includegraphics[scale=0.5,clip=true]{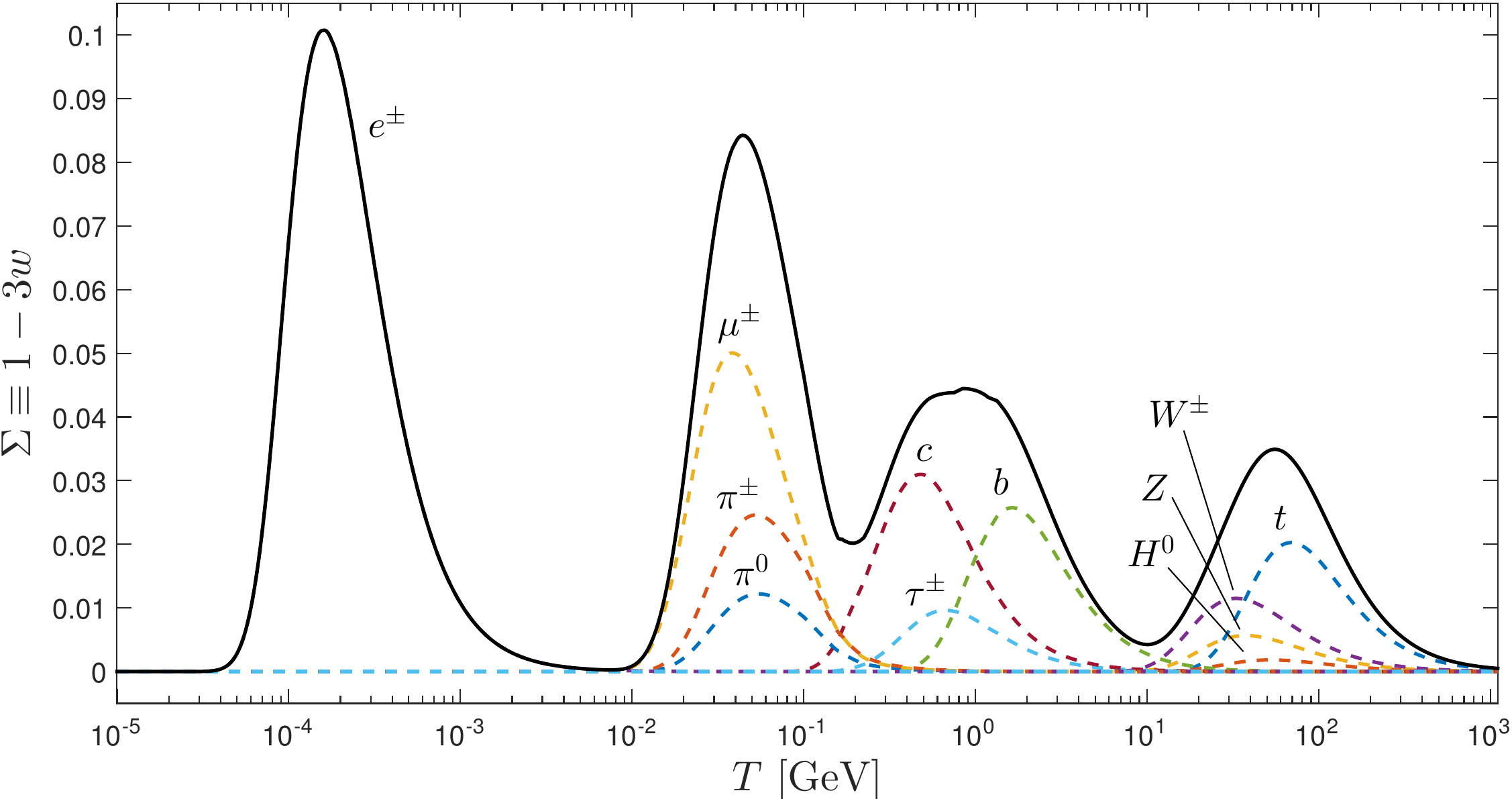}
\caption[Relativistic kick function $\Sigma(T)$]{Evolution of the quantity $\Sigma(T) \equiv 1 - 3w$ (solid black curve), 
defined in~\eqref{eq:sigma}, over the radiation era. The contributions from each particle species listed in table~\ref{tab:part_spec} 
are also shown (coloured dashed curves).}
\label{fig:sigma_rel}
\end{figure}

Using the Standard Model particle spectrum given in table~\ref{tab:part_spec}, we numerically evaluate~\eqref{eq:sigma}. The results for the contribution from each of the individual particle species (dashed curves) as well as their combined sum (solid black curve) are shown in 
figure~\ref{fig:sigma_rel}. The quantity $\Sigma(T)$ has been evaluated in several other publications (e.g.~\cite{PhysRevD.70.123518,PhysRevD.73.083525}) and in most cases the value of $g_{*\rho}(T)$ is assumed constant during the crossing of each particle threshold. However, since $g_{*\rho}(T)$ actually decreases during this interval, these calculations underestimate the height of $\Sigma(T)$, particularly for the final $e^{\pm}$ 'kick'. In our work we have maintained the temperature dependence of $g_{*\rho}(T)$ and our results agree with those contained in the recent paper by Erickcek~\textit{et al}~\cite{PhysRevLett.110.171101}.
\begin{table}
\begin{center}
\begin{tabular}[]{@{}c c c c }
\toprule
 & Particle  & Mass (GeV) & $g_i$ \\ \midrule
\multirow{10}{*}{Fermions} &$t$       & $173.07$    & $12$   \\
 &$b$       & $4.18$    & $12$   \\
 &$\tau^{\pm}$    & $1.78$    & $4$    \\
 &$c$       & $1.28$    & $12$   \\
 &$\mu^{\pm}$     & $0.106$   & $4$    \\
 &$s$       & $0.095$    & $12$   \\
 &$d$       & $4.8\times 10^{-3}$    & $12$   \\
 &$u$       & $2.3\times 10^{-3}$    & $12$   \\
 &$e^{\pm}$       & $5.11\times 10^{-4}$& 4  \\
 & $\nu$      & - & $6$  \\
\\
\multirow{7}{*}{Bosons} 
 & $H$      & $125.00$  & $1$    \\
 & $Z$      & $91.19$   & $3$    \\
 &$W^{\pm}$ & $80.39$   & $6$    \\
 &$^*\pi^{0}$   & $0.140$   & $1$    \\
 &$^*\pi^{\pm}$& $0.135$  & $2$    \\
 &$\gamma$ & - & $2$   \\ 
 &$g$ & - & $16$ 
          \\ \bottomrule \\
\end{tabular}
\caption[Properties of the Standard Model particles.]{\label{tab:part_spec} Properties of the Standard Model particle spectrum including mass (GeV) and number of spin degrees of freedom $g_i$ used in the evaluation of $\Sigma(T)$. Composite particles are marked with an asterisk.} 
\end{center}
\end{table}

\begin{figure}[htbp]
\centering
\includegraphics[scale=0.7,clip=true]{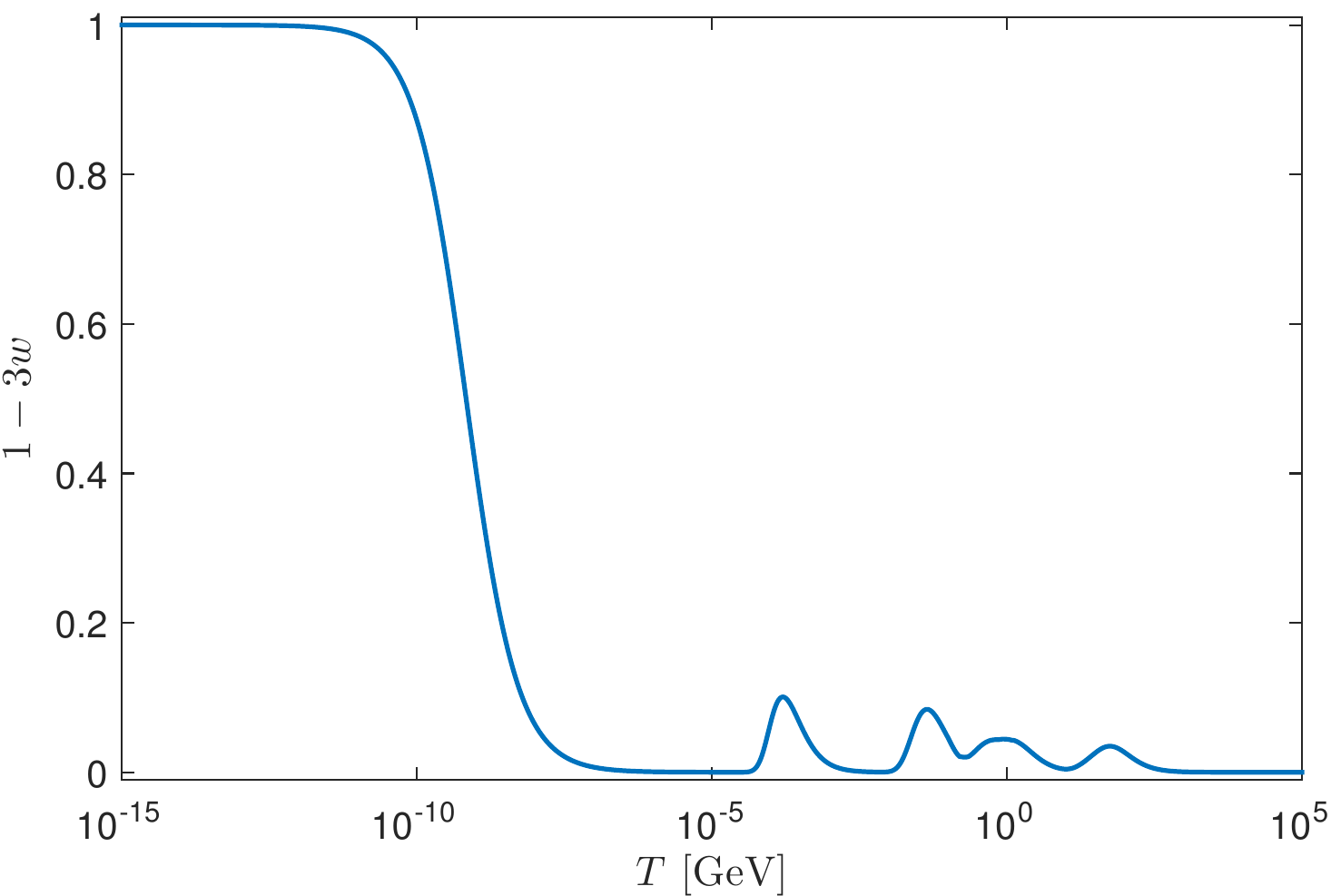}
\caption[Equation of state parameter]{Evolution of the quantity $\Sigma(T) \equiv 1 - 3w$ defined in~\eqref{eq:sigma} from the radiation era through to the matter era.}
\label{fig:Oneminus3w}
\end{figure}

After the final $e^{\pm}$ kick, the forcing term remains inactive until the universe transitions from the radiation dominated era to the matter dominated era in which
\begin{equation}
1 - 3w \simeq \frac{\rho_{\mathrm{m}}}{\rho_{\mathrm{m}} + \rho_{\mathrm{r}}} \simeq \frac{1}{1 + T/T_{\mathrm{eq}}}\label{eq:sigma_matdom}
\end{equation}
where $T_{\mathrm{eq}}\sim \mathcal{O}(10^{-9})$ GeV is the temperature of matter-radiation equality, i.e. $\rho_{\mathrm{r}}(T_{\mathrm{eq}}) = \rho_{\mathrm{m}}(T_{\mathrm{eq}})$. Combining the results shown in figure~\ref{fig:sigma_rel} with the late time behaviour described by equation~\eqref{eq:sigma_matdom}, we obtain the evolution of $1 - 3w$ from the radiation era up to the present time. This is shown in figure~\ref{fig:Oneminus3w} 
where we observe the general features that, in the radiation era, $T \gtrsim T_{\mathrm{eq}}$, $1 - 3w$ is $\approx 0$ except at the location of each of the kicks; then, as $T$ approaches $T_{\mathrm{eq}}$, $1 -3w$ smoothly rises to $1/2$ and, in the limit $T \ll T_{\mathrm{eq}}$, $1 - 3w$ is $\approx 1$. 

\begin{figure}[htbp]
\centering
\includegraphics[scale=0.8,clip=true]{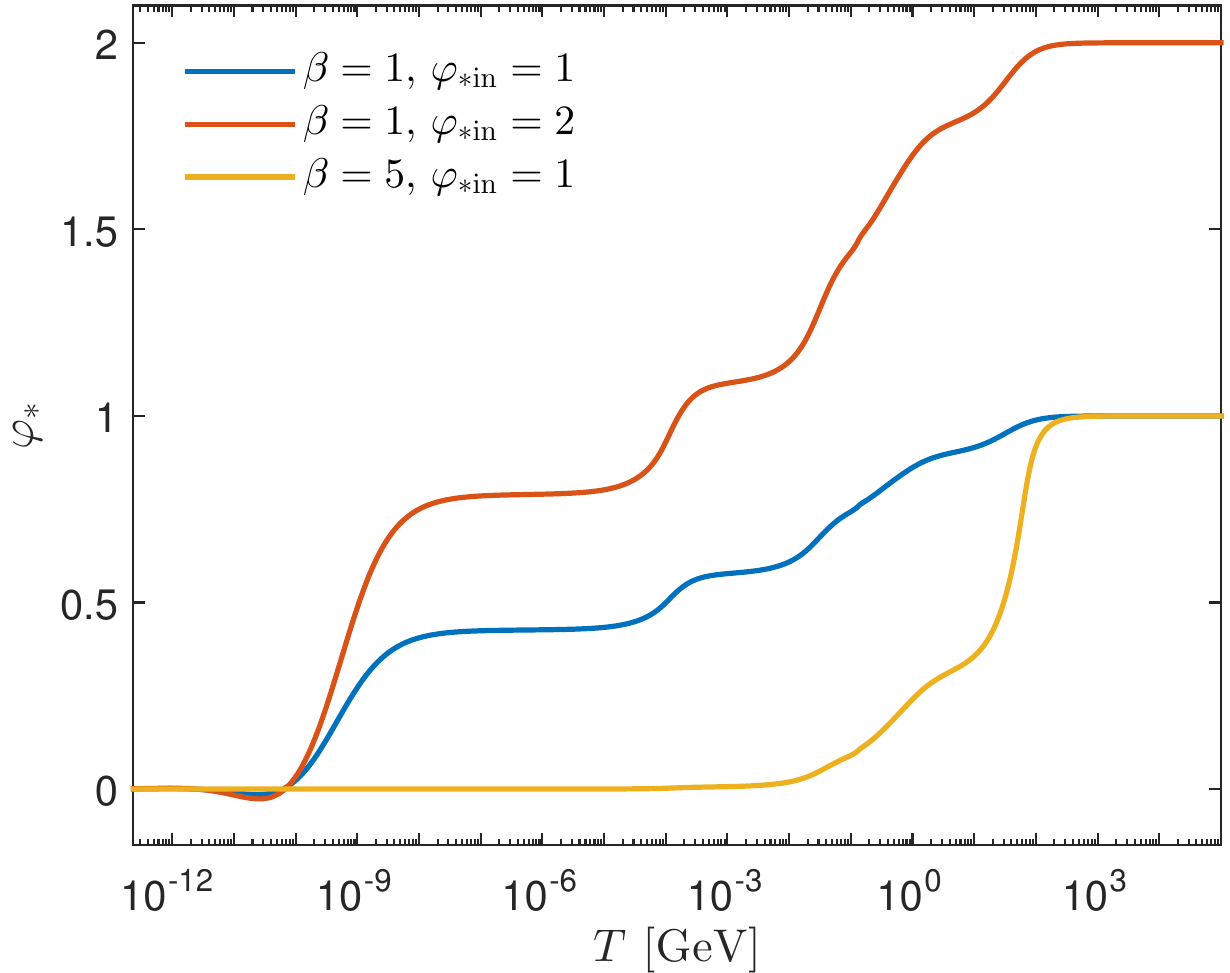}
\caption[Scalar field evolution in scalar-tensor gravity models]{Evolution of $\varphi_*$ for different values of the input parameters $\beta$ and $\varphi_{*\mathrm{in}}$.}
\label{fig:phi_evolution}
\end{figure}

The cosmological evolution of the scalar field $\varphi_*$ can now be determined from \eqref{eq:phimotion_universal}. Substituting the computed values of $1 - 3w$ and our particular choice of coupling $A(\varphi_*) = e^{\frac{1}{2}\beta\varphi_*^2}$ into~\eqref{eq:phimotion_universal}, we numerically integrate the equation of motion for several sample values of $\beta$ and $\varphi_{*\mathrm{in}}$ (see figure~\ref{fig:phi_evolution}). 

In each case we see that the field is attracted towards $\varphi_* = 0$ which, for the quadratic coupling considered, corresponds to the GR limit. Importantly we notice that the attraction mechanism is initiated well before the radiation-matter transition at $T\sim 10^{-9}$ GeV due to the non-relativistic kicks mentioned above. In fact, for the $\beta = 1$ cases (blue and red curves) we can discern the four distinct kicks corresponding to the four peaks in figure~\ref{fig:sigma_rel}. For the $(\beta,\varphi_{*\mathrm{in}})=(5,1)$ case (yellow curve), the attraction mechanism is so efficient that the field approaches the GR limit at much earlier times, prior to BBN.


\section{Modified expansion rate}
\label{sec:st_modhub}

Having calculated the cosmological evolution of $\varphi_*$ we can now determine the modified expansion rate in the scalar-tensor scenario. For the purpose of calculating the dark matter relic abundance, we are particularly interested in the Jordan frame expansion rate $H = d\ln a/dt$ since the dark matter particles couple directly to the Jordan frame metric $g_{\mu\nu}$ and in this frame particle masses, number densities, etc. take their standard form. Therefore the expansion rate that governs the timing of particle decoupling and in turn the dark matter relic abundance is determined using~\eqref{eq:hjord}:
\begin{equation}
H = A^{-1}(\varphi_*)H_*\left[1 + \alpha(\varphi_*)\varphi_*'\right]\label{eq:hjord2}
\end{equation}
where, using~\eqref{eq:ein_hub_p} and~\eqref{eq:einjordrho}, we can write the Einstein frame expansion rate, $H_*$, as
\begin{equation}
H_*^2 = \frac{8\pi G_*}{3 - \varphi_*'^2}\,\rho A^4(\varphi_*).
\end{equation}
To compare the Jordan frame expansion rate, $H$, to the expansion rate in the standard cosmological scenario,
\begin{equation}
H_{\mathrm{GR}}^2 = \frac{8\pi G}{3}\rho,
\end{equation}
we note that the gravitational couplings used in each case are related via~\cite{1970ApJ...161.1059N}
\begin{equation}
G = G_* A^2({\varphi_{*0}})\left[1 + \alpha^2(\varphi_{*0})\right],
\end{equation}
where $G$ is the gravitational coupling given in the standard General Relativity scenario and $G_*$ is the bare gravitational coupling used here. Combining this expression with the connection between the Jordan and Einstein frame Hubble factors~\eqref{eq:hjord} we finally have
\begin{equation}
\xi \equiv \frac{H_{\mathrm{ST}}}{H_{\mathrm{GR}}} = \frac{A(\varphi_*)}{A(\varphi_{*0})}\frac{1 + \alpha(\varphi_*)\varphi_*'}{\sqrt{1 - \varphi_*'^2/3}}\frac{1}{\sqrt{1 + \alpha^2(\varphi_{*0})}}\label{eq:xi_stgrav}
\end{equation}
where we now denote the Jordan frame expansion rate $H$, that will be used as input into the Boltzmann rate equation for the dark matter number density, by $H_{\mathrm{ST}}$.

\begin{figure}[tbp]
\centering 
\includegraphics[scale=0.8,trim=0 0 0 0,clip=true]{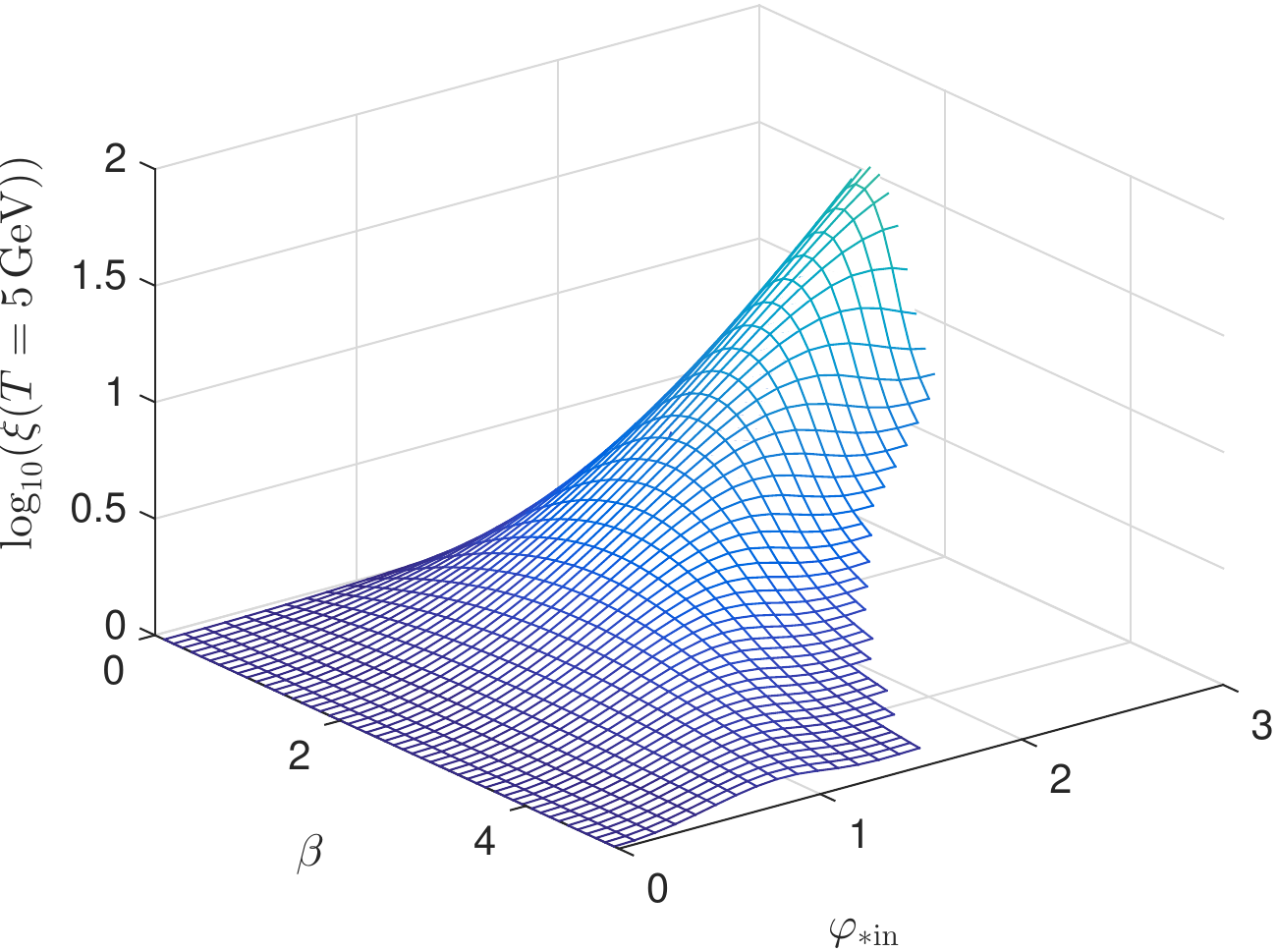}
\caption[Ratio of the scalar-tensor and standard expansion rates at the time of dark matter decoupling]{\label{fig:dilaton_xi_fo} Magnitude of the ratio of the scalar-tensor and standard expansion rates, $\xi \equiv H_{\mathrm{ST}}/H_{\mathrm{GR}}$, evaluated at a temperature $T = 5$ GeV as a function of the input parameters $\beta$ and $\varphi_{*\mathrm{in}}$, which satisfy the dynamical constraints.}
\end{figure}

To estimate the possible enhancement of the dark matter relic abundance in the scalar-tensor gravity scenario, we first determine the magnitude of the ratio $\xi = H_{\mathrm{ST}}/H_{\mathrm{GR}}$ around the time of dark matter decoupling. Hence, in figure~\ref{fig:dilaton_xi_fo}, we plot the magnitude of $\xi$ evaluated at $T_{f} \sim m_\chi/20 \sim 5$ GeV, which is the typical freeze-out time for a $100$ GeV WIMP. We have chosen values of $\beta > 0$ to ensure that the model is dynamically attracted towards GR (see previous section) and omitted data points for larger $\beta$ and $\varphi_{*\mathrm{in}}$ that violate the dynamical constraints, leading to unphysical results (see next section).

This figure shows that the scalar-tensor expansion rate at the time of dark matter decoupling can be more than an order of magnitude larger than the standard expansion rate. In particular, the largest enhancements are seen for small values of $\beta$ and large $\varphi_{*\mathrm{in}}$, specifically, for $\beta \lesssim 2$ and $\varphi_{*\mathrm{in}}\gtrsim 2$. This parameter range leads to large initial values of $\xi_{\mathrm{in}}\sim A(\varphi_{*\mathrm{in}})$ (because of the large $\varphi_{*\mathrm{in}}$) whilst exhibiting a less efficient attraction towards GR (small $\beta$) so that the expansion rate at the time of freeze-out still deviates significantly from the standard one. In general, although increasing both $\beta$ and $\varphi_{*\mathrm{in}}$ increases the initial value $\xi_{\mathrm{in}}$, because the displacement of the field towards $\varphi_* = 0$ due to each of the non-relativistic 'kicks' also increases with increasing $\beta$ and $\varphi_{*\mathrm{in}}$, the overall effect on $\xi(T_f)$ (and in turn on $\Omega_{\mathrm{DM}}^{\mathrm{ST}}/\Omega_{\mathrm{DM}}^{\mathrm{GR}}$) can be difficult to predict. 

Although the early time expansion rate can certainly be much larger in the scalar-tensor scenario for certain regions of parameter space, we must be careful to ensure that such regions satisfy both the astrophysical and dynamical constraints placed on the scalar field and its evolution. In the next section we investigate in more detail these constraints and determine those points in parameter space which are viable.


\section{Constraints}
\label{sec:st_constraints}

Solar system tests of gravity, including the perihelion shift of Mercury and Lunar Laser Ranging experiments, place strict constraints on deviations from General Relativity (see for example~\cite{Will:2014xja}). Most relevant for our purposes are the measurements of the Shapiro time delay performed by the Cassini spacecraft~\cite{Cassini} which indicate that the present value of the scalar-tensor deviation parameter $\alpha_0^2\equiv \alpha^2(\varphi_{*0}) < 10^{-5}$. Hence, if the gravitational interaction is truly described by a scalar-tensor theory, it must be extremely close to General Relativity in our local neighbourhood.\footnote{Additionally, for the quadratic scalar coupling $A(\varphi_*) = e^{\frac{1}{2}\beta\varphi_*^2}$ considered here, the decay of the orbital period of pulsars in asymmetric binaries implies that the coupling parameter $\beta \gtrsim -4.5$~\cite{freire2012relativistic}.}

\begin{figure}[htbp]
\centering
\includegraphics[scale=0.8,clip=true]{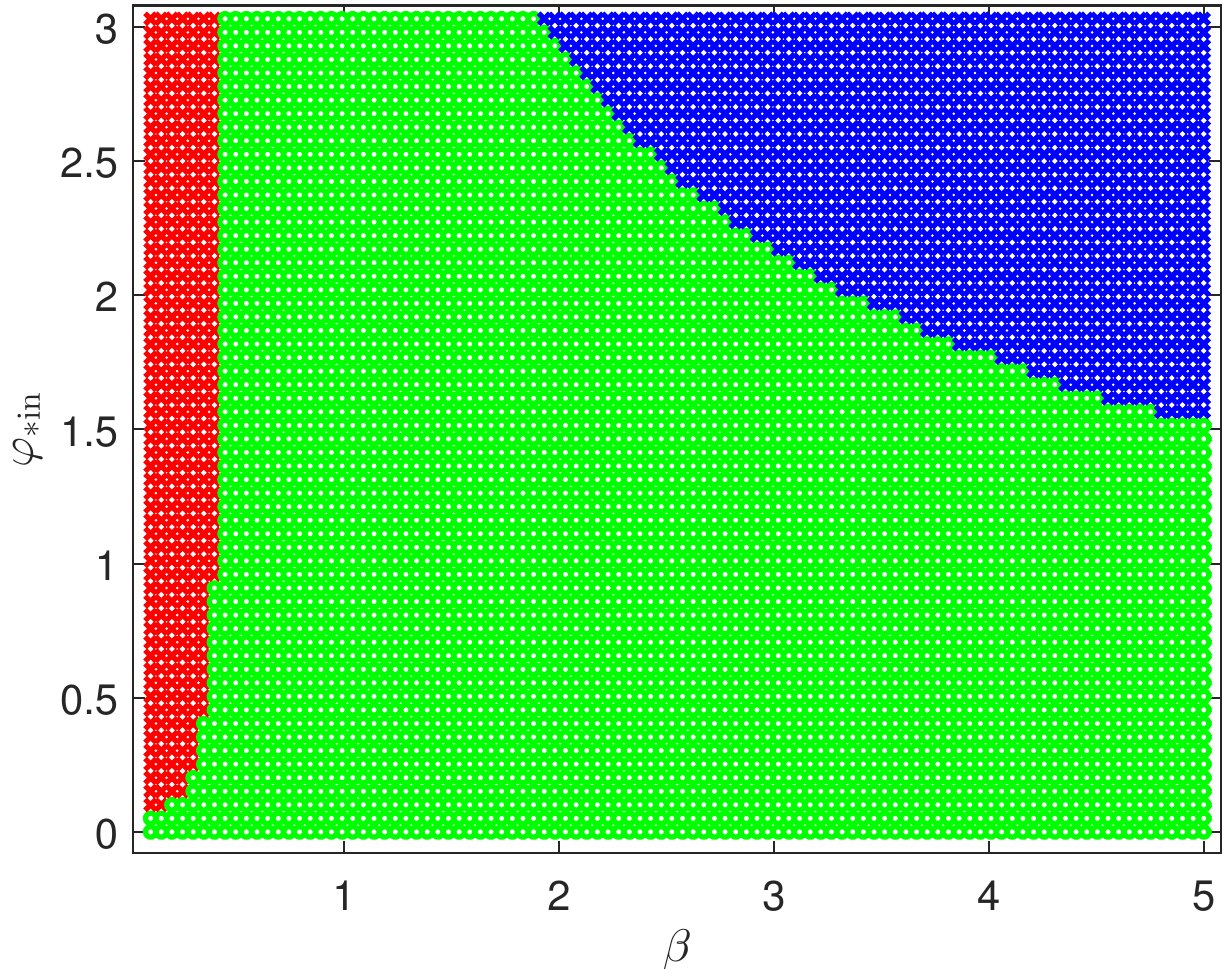}
\caption[Allowed region in the ($\beta,\varphi_{*\mathrm{in}}$) parameter space satisfying the Cassini bounds]{Exclusion plot following a scan of the input parameter space $(\beta,\varphi_{*\mathrm{in}})$ and applying the Cassini bound $\alpha_0^2 < 10^{-5}$~\cite{Cassini}. The points excluded due to the Cassini (dynamical) constraints have been marked with a red (blue) cross and the allowed points are indicated by the green circles.}
\label{fig:exclusion_zone_cassini}
\end{figure}

Although this constraint only applies to the present value $|\alpha_0|$, we can relate it to the input parameters $\beta$ and $\varphi_{*\mathrm{in}}$ by integrating the $\varphi_*$ evolution equation up to the present epoch and calculating the predicted value of $|\alpha_0|$ directly. We can then determine which values of $\beta$ and $\varphi_{*\mathrm{in}}$ satisfy the Cassini bound $\alpha_0^2 \lesssim 10^{-5}$. The results are shown in figure~\ref{fig:exclusion_zone_cassini} where we have indicated those points that violate the Cassini bound with a red cross and those that are acceptable with a green circle. We have also indicated with a blue cross those points that violate the various dynamical constraints on the evolution of the scalar field (see later). 

Interestingly, for the parameter range considered, the Cassini bound only excludes values of $\beta \lesssim 0.4$. Comparing this with the results in figure~\ref{fig:dilaton_xi_fo} for $\xi(T_f)$, we see that the regions of parameter space that lead to the fastest expansion rates at the time of dark matter decoupling $(0.4\lesssim \beta\lesssim 2, \varphi_{*\mathrm{in}} \gtrsim 2)$ still satisfy the constraint $\alpha_0^2<10^{-5}$. Therefore, the Cassini bound alone (or, more generally, solar system tests of gravity) does not preclude large deviations from the standard cosmological history at early times. This is not so surprising given that solar system tests of gravity rely on data taken at late times --- \textit{long} after the attraction mechanism towards GR has been initiated.

To properly constrain the input parameters and the behaviour of the field in the early universe, we must also take into account the results presented in Coc~\textit{et al}~\cite{PhysRevD.73.083525} where they considered the Big Bang Nucleosynthesis implications of scalar-tensor theories with a quadratic coupling, i.e. $A(\varphi_*) = e^{\frac{1}{2}\beta\varphi_*^2}$. In this work the authors perform a full numerical integration of the scalar field evolution equation and calculate the resulting light element abundances using an up-to-date BBN code. The results of these calculations are then compared to observed light element abundances to constrain the various scalar-tensor model parameters.

\begin{figure}[htbp]
\centering
\includegraphics[scale=0.8,clip=true]{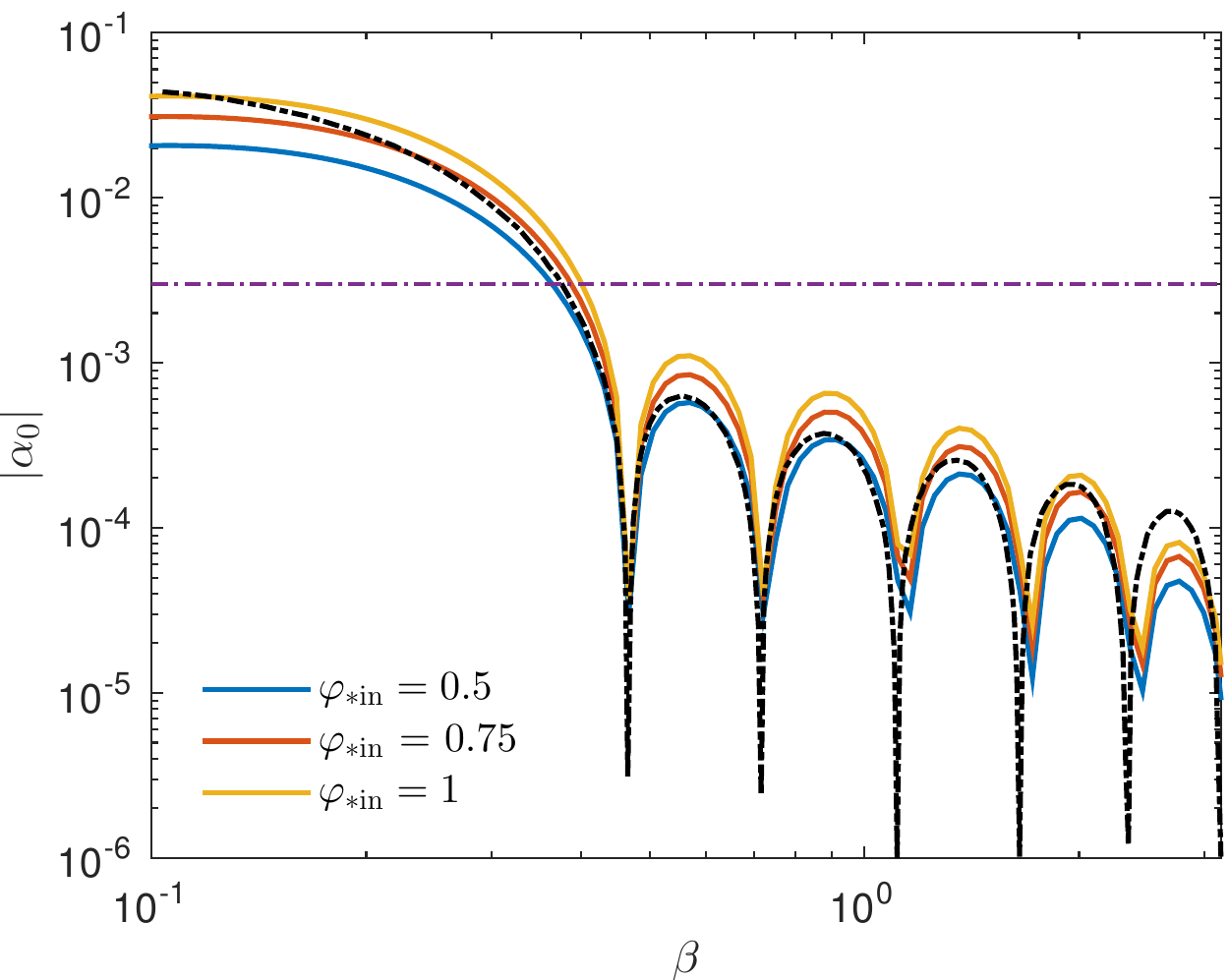}
\caption[Comparison of the scalar-tensor predictions with Big-Bang Nucleosynthesis constraints]{Magnitude of $|\alpha_0|$ as a function of the coupling parameter $\beta$ for different initial values of the field (solid curves). The results are compared with the constraints derived using the solar system tests of gravity (dot-dashed purple) and the BBN constraints given in figure 19 of Coc \textit{et al}~\cite{PhysRevD.73.083525} (dot-dashed black).}
\label{fig:bbn_constraints_st}
\end{figure}

Since these constraints are also given in terms of the present value $|\alpha_0|$, we must again integrate the $\varphi_*$ evolution equation up to the present epoch for the different values of $\beta$ and $\varphi_{*\mathrm{in}}$ and calculate $|\alpha_0|$ directly.
As an example, in figure~\ref{fig:bbn_constraints_st}, we compare our results for $|\alpha_0|$ as a function of $\beta$ for different starting values of $\varphi_{*\mathrm{in}}$ (solid curves) with both the Cassini (dot-dashed purple curve) and the Coc~\textit{et al} BBN bound (dot-dashed black curve). 

\begin{figure}[htbp]
\centering
\includegraphics[scale=0.8,clip=true]{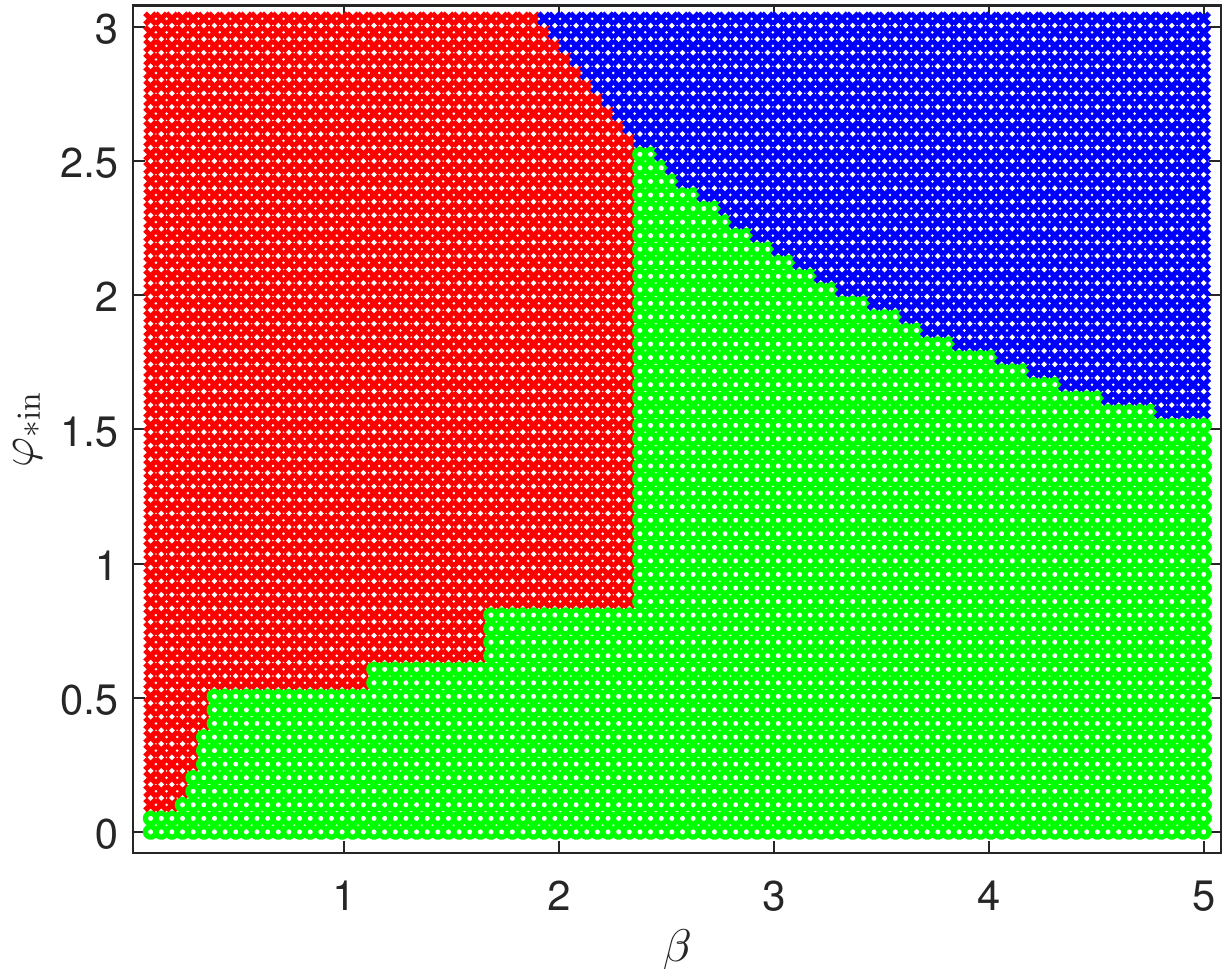}
\caption[Allowed region in the $(\beta,\varphi_{*\mathrm{in}}$) parameter space satisfying the Cassini bounds and the bounds from BBN calculations]{Scan of the input parameter space for the coupling $A(\varphi_*) = e^{\frac{1}{2}\beta\varphi_*^2}$. The points excluded due to astrophysical (dynamical) constraints have been marked with a red (blue) cross and the allowed points are indicated by the green circles.}
\label{fig:dilaton_exclusion_zone}
\end{figure}

Ignoring any computational differences, (keep in mind that our computation of $\Sigma(T)$ differs from figure 5 of Coc~\textit{et al}~\cite{PhysRevD.73.083525} because we maintained the temperature dependence of $g_{*\rho}(T)$) we notice that the blue, red and yellow curves for $|\alpha_0|$ do not drop below the BBN constraints (dot-dashed black curve) until after a particular oscillation. For example, for $\varphi_{*\mathrm{in}} = 0.75$ (red curve), the calculated values of $|\alpha_0|$ do not drop below the BBN constraints until after the fourth oscillation at $\beta \simeq 1.65$. Similarly, for $\varphi_{*\mathrm{in}} = 1$ (yellow curve), $|\alpha_0|$ does not drop below the BBN bound until after the fifth oscillation at $\beta = 2.35$. Therefore, once the BBN bounds become more stringent than the solar system bound, i.e. $\beta \gtrsim 0.4$, the values of $\beta$ for which a particular $\varphi_{*\mathrm{in}}$ becomes acceptable are discrete. This is shown in figure~\ref{fig:dilaton_exclusion_zone} where we reproduce figure~\ref{fig:exclusion_zone_cassini}, this time applying the Coc~\textit{et al} BBN bound in addition to the Cassini bound. For $\beta \lesssim 0.4$, the boundary separating the allowed and rejected regions follows a smooth curve, whilst for $\beta \gtrsim 0.4$ the boundary increases in discrete steps.

Most importantly, a comparison of figures~\ref{fig:exclusion_zone_cassini} and~\ref{fig:dilaton_exclusion_zone} shows that applying the Coc~\textit{et al} BBN constraints excludes most of the parameter space corresponding to large early time expansion rates (see figure~\ref{fig:dilaton_xi_fo}).
Also, because of the non-injective relationship between the input parameters and the present value of $|\alpha_0|$ (i.e. different initial conditions can lead to the same $|\alpha_0|$), the constraints in figure 19 of~\cite{PhysRevD.73.083525} can only be treated as a strict upper bound. This means that values of $|\alpha_0|$ lower than this upper bound are not necessarily acceptable so that the allowed region in the ($\beta,\varphi_{*\mathrm{in}}$) parameter space is most likely smaller than what is shown in figure~\ref{fig:dilaton_exclusion_zone}.

In addition to the astrophysical constraints, the dynamical evolution of the scalar field $\varphi_*$ must also satisfy the condition
\begin{equation}
1 + \alpha(\varphi_*)\varphi_*' > 0.
\end{equation}
This ensures that the Jordan frame Hubble factor, derived from the Einstein frame Hubble factor through
\begin{equation}
H = A^{-1}(\varphi_*)H_*\left[1 + \alpha(\varphi_*)\varphi_*'\right],
\end{equation}
is positive definite. Moreover, if this condition is violated, the Jordan frame temperature $T$ becomes a multi-valued function of the evolution parameter $N$ and the conformal transformation between the Jordan and Einstein frames breaks down.\footnote{If we neglect the temperature dependence of $g_{*s}(T)$, the connection between the Jordan frame temperature and the evolution parameter $N$~\eqref{eq:TtoN} gives
\begin{equation}
\frac{dT}{dN} = -T_0\frac{A(\varphi_{*0})}{A(\varphi_*)}e^{-N}\left[1 + \alpha(\varphi_*)\varphi_*'\right].
\end{equation}
If the quantity $1 + \alpha(\varphi_*)\varphi_*'$ changes sign, the relationship between $T$ and $N$ is not monotonic and $T$ becomes a multi-valued function of $N$ that cannot be inverted.} The points in the $(\beta,\varphi_{*\mathrm{in}})$ plane that violate this dynamical constraint are indicated with blue crosses.

We can crudely estimate the general shape of the dynamical exclusion zone by approximating the maximum velocity reached by the field due to each 'kick' as $\varphi_{*{\mathrm{max}}}' \sim \alpha(\varphi_*)$. The condition $1 + \alpha(\varphi_*)\varphi_*'>0$ then becomes
\begin{equation}
\alpha(\varphi_*)^2 \lesssim \mathcal{D}^2,
\end{equation}
where $\mathcal{D}$ is some variable that depends on the magnitude of the kick. Substituting in the expression $\alpha(\varphi_*) = \beta\varphi_*$, we find that the values of $\varphi_*$ in the region
\begin{equation}
\varphi_* \gtrsim \frac{\mathcal{D}}{\beta}\,,\\
\end{equation}
are excluded. 

Combining the astrophysical and dynamical constraints we see that a significant portion of the parameter space is excluded. In particular, the region corresponding to the greatest deviations from the standard expansion history at the time of dark matter decoupling (i.e. $\xi(T_f)$) is no longer allowed (compare figures~\ref{fig:dilaton_exclusion_zone} and~\ref{fig:dilaton_xi_fo}). In fact the largest value of $\xi(T_f)$ permitted by the various constraints is only $\sim 4$. Hence, although large early-time expansion rates can be achieved within scalar-tensor cosmological models with the coupling $A(\varphi_*) = e^{\frac{1}{2}\beta\varphi_*^2}$, those models that satisfy the various constraints mentioned above will not significantly deviate from the standard cosmological model at the time of dark matter decoupling. In turn, we expect that the relic abundance of dark matter to be relatively unaffected and that any enhancements with respect to the canonical result are modest. In the next section we verify this assertion by numerically solving the Boltzmann rate equation governing the cosmic evolution of the dark matter number density to determine the present dark matter density.


\section{Symmetric dark matter}
\label{sec:st_sdm}

The relic abundance of a symmetric dark matter species $\chi (= \bar{\chi})$, initially in equilibrium with the background cosmic bath, is determined from the dark matter number density $n_\chi$ which evolves according to the relativistic Boltzmann equation
\begin{equation}
\frac{dn_\chi}{dt} = -3Hn_\chi - \langle\sigma v\rangle\left(n_\chi^2 - n_\chi^{\mathrm{eq}\,^2}\right).\label{eq:boltsym}
\end{equation}
Here $n_\chi^{\mathrm{eq}}$ is the equilibrium number density and $\langle\sigma v\rangle$ is the thermally averaged annihilation cross section times relative velocity (loosely termed the "annihilation cross section").  We adopt the generic parameterization for the annihilation cross section~\cite{PhysRevD.33.1585,PhysRevLett.64.615}
\begin{equation}
\langle\sigma v\rangle = a + \frac{bT}{m_\chi},
\end{equation}
where the constant term, $a$, corresponds to $s-$wave scattering and the temperature dependent term, $b$, to $p-$wave scattering.

The Boltzmann equation~\eqref{eq:boltsym} can be rewritten in terms of $x = m_\chi/T$ and the comoving number density $Y=n_\chi/s$, where $s$ is the entropy density given by $s = 2\pi^2 g_*(T) T^4/45$.\footnote{Here $g_*(T)$ actually refers to the number of entropic degrees of freedom $g_{* s}$. Since the number of relativistic and entropic degrees of freedom only differ when a particle crosses a mass threshold, we take $g_{*\rho}=g_{* s} \equiv g_{*}$~\cite{Steigman:2012nb}.} 
We then have
\begin{equation}
\frac{dY}{dx}=-\frac{s\langle\sigma v\rangle}{xH} \zeta(x)\left(Y^2 - Y_{\mathrm{eq}}^2\right),~\label{eq:bolt}
\end{equation}
where $Y_{\mathrm{eq}} \simeq 0.145(g_\chi/g_*)\,x^{3/2}e^{-x}$, $g_\chi = 2$ is the number of internal degrees of freedom of the dark matter species $\chi$ and
\begin{equation} 
\zeta (x) = 1 - \frac{1}{3}\frac{d\log{g_*}}{d\log{x}}~\label{eq:df} 
\end{equation}
is a temperature dependent factor related to the change in the number of degrees of freedom.
The present dark matter density, $\Omega_{\mathrm{DM}}h^2$, is obtained from the asymptotic 
solution ($x\rightarrow\infty$) of equation~\eqref{eq:bolt}
\begin{equation}
\Omega_{\mathrm{DM}}h^2 = 2.75\times 10^8\,m_\chi Y_\infty,\label{eq:omgdm}
\end{equation}
where $Y_\infty=Y(x \rightarrow \infty)$ is the present comoving density. 

As discussed in section~\ref{sec:conformal_frames}, each of the quantities in~\eqref{eq:bolt} such as the comoving number density, $Y = n/s$, and annihilation cross section, $\langle\sigma v\rangle$, is defined in the Jordan frame where particle physics quantities take their standard interpretation. Hence $H$ is the Jordan frame expansion rate determined using~\eqref{eq:hjord}.\footnote{In the universal coupling case there is a unique Jordan frame in which both the Standard Model particles and the dark matter particles couple directly to the metric $g_{\mu\nu}$. This situation becomes more complicated in non-universal scalar-tensor theories where the coupling with the \textit{visible} and \textit{dark} sectors is distinct (see section~\ref{sec:st_nonuniversal}).} 

Taking the results from our scan over the input parameter space, we numerically integrate the $\varphi_*$ evolution equation~\eqref{eq:phimotion_universal} and determine the Jordan frame expansion rate $H$ from~\eqref{eq:hjord2}. In doing so we have chosen values of $\beta$ and $\varphi_{*\mathrm{in}}$ that are (i) permitted by the astrophysical and dynamical constraints (according to figure~\ref{fig:dilaton_exclusion_zone}) and (ii) provide the greatest deviation from the standard expansion history (see figure~\ref{fig:dilaton_xi_fo}). 

\begin{figure}[tbp]
\centering
\includegraphics[scale=0.53,trim=0 0 0 0,clip=true]{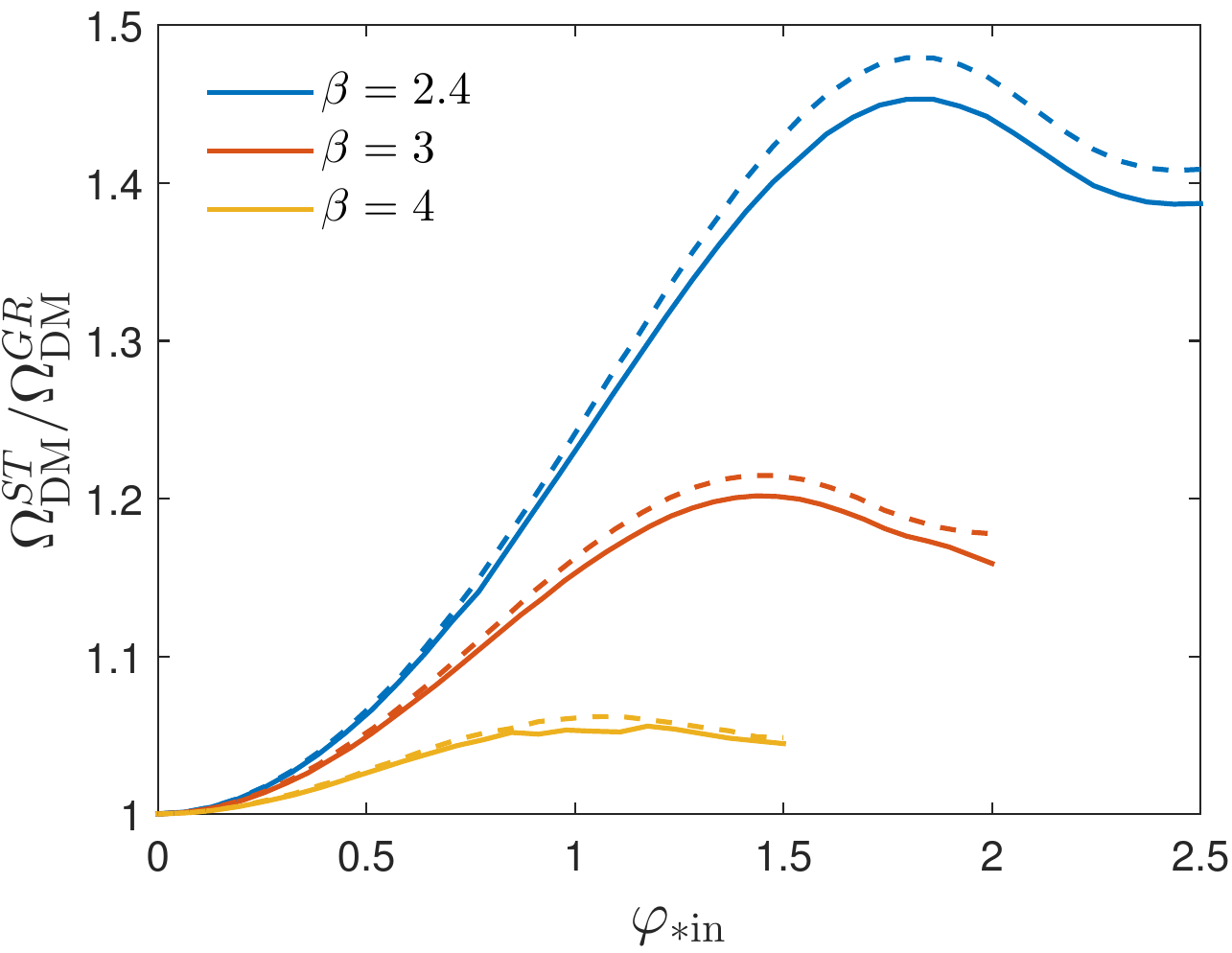}
\hfill
\includegraphics[scale=0.53,trim=0 0 0 0,clip=true]{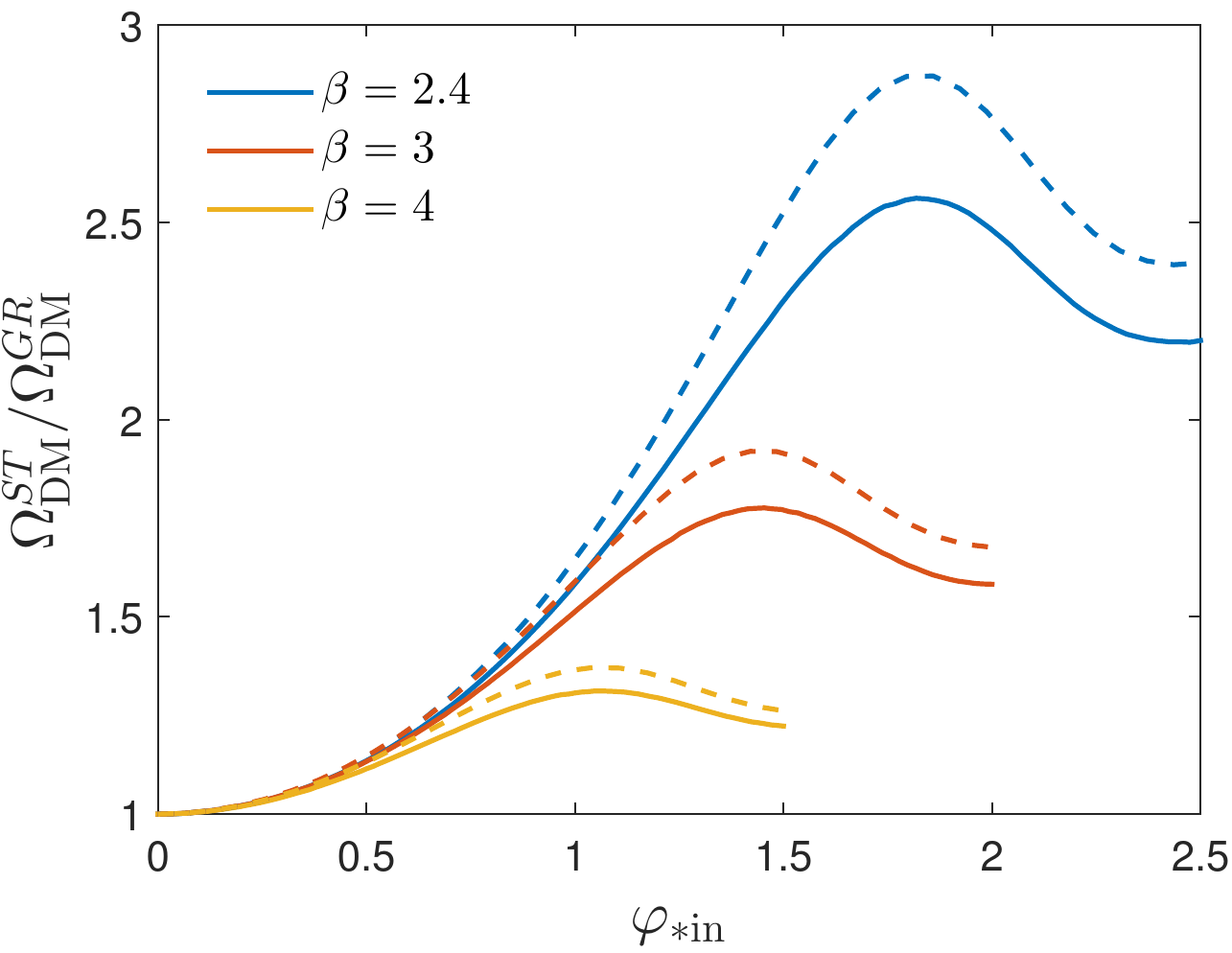}
\caption[Dark matter relic abundance enhancement in scalar-tensor gravity models]{\label{fig:coc_universal_omega} Ratio of the relic abundance of symmetric dark matter in the scalar-tensor and standard cosmological scenarios for $m_\chi = 10$ GeV (left panel) and $m_\chi = 100$ GeV (right panel) as a function of $\varphi_{*{\mathrm{in}}}$ for different values of $\beta$. The solid (dashed) curves correspond to $s(p)-$wave annihilation.}
\end{figure}

Substituting this result into~\eqref{eq:bolt} we solve the Boltzmann equation to determine the present dark matter density. The results are shown in figure~\ref{fig:coc_universal_omega} where we plot the ratio of the dark matter relic abundance in the scalar-tensor gravity model, $\Omega_{\mathrm{DM}}^{\mathrm{ST}}$, to the corresponding value in the standard cosmological scenario, $\Omega_{\mathrm{DM}}^{\mathrm{GR}}$ (i.e. the solution to~\eqref{eq:bolt} using $H_{\mathrm{GR}}$), as a function of $\varphi_{*\mathrm{in}}$ for different values of $\beta$. In the left and right panels of figure~\ref{fig:coc_universal_omega} we have calculated the relic abundance enhancement ratio $\Omega_{\mathrm{DM}}^{\mathrm{ST}}/\Omega_{\mathrm{DM}}^{\mathrm{GR}}$ for $m_\chi = 10$ GeV and $m_\chi = 100$ GeV respectively with the solid (dashed) curves corresponding to $s$($p$)$-$wave annihilation. Note that for the values of $\beta$ chosen, namely $\beta = (2.4,3.0,4.0)$, we have only calculated results up to $\varphi_{*\mathrm{in}} = (2.5,2.0,1.5)$ respectively because larger values of $\varphi_{*\mathrm{in}}$ are excluded due to dynamical constraints (see figure~\ref{fig:dilaton_exclusion_zone}).  

In each panel we see that the ratio $\Omega_{\mathrm{DM}}^{\mathrm{ST}}/\Omega_{\mathrm{DM}}^{\mathrm{GR}}$ increases for small $\varphi_{*\mathrm{in}}$ reaching a $\beta$-dependent maximum value before falling as $\varphi_{*\mathrm{in}}$ increases further. For the $m_\chi = 10$ GeV and $m_\chi = 100$ GeV cases the maximum enhancement for $s-$wave annihilation is only $\Omega_{\mathrm{DM}}^{\mathrm{ST}}/\Omega_{\mathrm{DM}}^{\mathrm{GR}} \sim 1.5$ and $2.6$ respectively and is obtained for $(\beta,\varphi_{\mathrm{in}}) = (2.4,1.8)$ in both cases. For $p-$wave scattering and $m_\chi = 100$ GeV, this ratio increases to $\Omega_{\mathrm{DM}}^{\mathrm{ST}}/\Omega_{\mathrm{DM}}^{\mathrm{GR}}\sim 2.9$.

Because the lack of an algebraic expression for the Jordan frame expansion rate $H$ (see~\eqref{eq:xi_stgrav}) inhibits our ability to derive an approximate analytical solution for the dark matter relic density, it is difficult to make any quantitative estimates for the ratio $\Omega_{\mathrm{DM}}^{\mathrm{ST}}/\Omega_{\mathrm{DM}}^{\mathrm{GR}}$ in terms of $\beta$ and $\varphi_{*\mathrm{in}}$. Although we expect that the relic density enhancement factor, $\Omega_{\mathrm{DM}}^{\mathrm{ST}}/\Omega_{\mathrm{DM}}^{\mathrm{GR}}$, is related to the magnitude of $\xi = H_{\mathrm{ST}}/H_{\mathrm{GR}}$ evaluated at the time of dark matter decoupling, i.e. $\xi(T_f)$, the situation is still not straightforward given that the relationship between $\xi(T_f)$ and the input parameters $\beta$ and $\varphi_{*\mathrm{in}}$ is difficult to predict (see the discussion in section~\ref{sec:st_modhub}). We simply comment that the relic abundance curves displayed in figure~\ref{fig:coc_universal_omega} follow the same profile as the corresponding $\xi(T_f)$ curves evaluated at fixed $\varphi_{*\mathrm{in}}$. 

Most importantly, for the quadratic coupling $A(\varphi_*) = e^{\frac{1}{2}\beta\varphi_*^2}$ considered here, the maximum enhancement achieved for the $s-$wave case is only $\sim 2.6$ which, for the equivalent WIMP mass, is much less than the factor of $\sim 400$ reported by Catena~\textit{et al}~\cite{PhysRevD.70.063519}. Similarly, the maximum enhancement for the $p-$wave case, $\sim 2.9$, is several orders of magnitude below the factor of $\sim 1400$ given in~\cite{PhysRevD.70.063519}.

Moreover, we find no evidence for reannihilation. That is, during our scan over the input parameters we did not observe a secondary phase of dark matter annihilation following particle freeze-out. In the Catena~\textit{et al} 2004 paper~\cite{PhysRevD.70.063519}, reannihilation was a consequence of the rapid relaxation of the scalar-tensor expansion rate towards the standard expansion rate, which they purport occurs~\textit{after} dark matter decoupling. However, we showed in section~\ref{sec:scalar_dynamics} that the attraction mechanism towards General Relativity is initiated when the temperature of the universe first drops below the rest masses of Standard Model particles at $T \gtrsim 10^2$ GeV. Even taking the most optimistic estimate that $x_f \sim 10$ in the scalar-tensor scenario, a dark matter particle with $m_\chi = 50$ GeV (the same value used in figure 7 of~\cite{PhysRevD.70.063519}) would freeze out at $T_f = 5$ GeV. Therefore, by the time the dark matter particles decouple from the thermal background the attraction mechanism is typically well underway and the scalar-tensor expansion rate is already relatively close to the standard result.



\begin{figure}[tbp]
\centering
\includegraphics[scale=0.53,trim=0 0 0 0,clip=true]{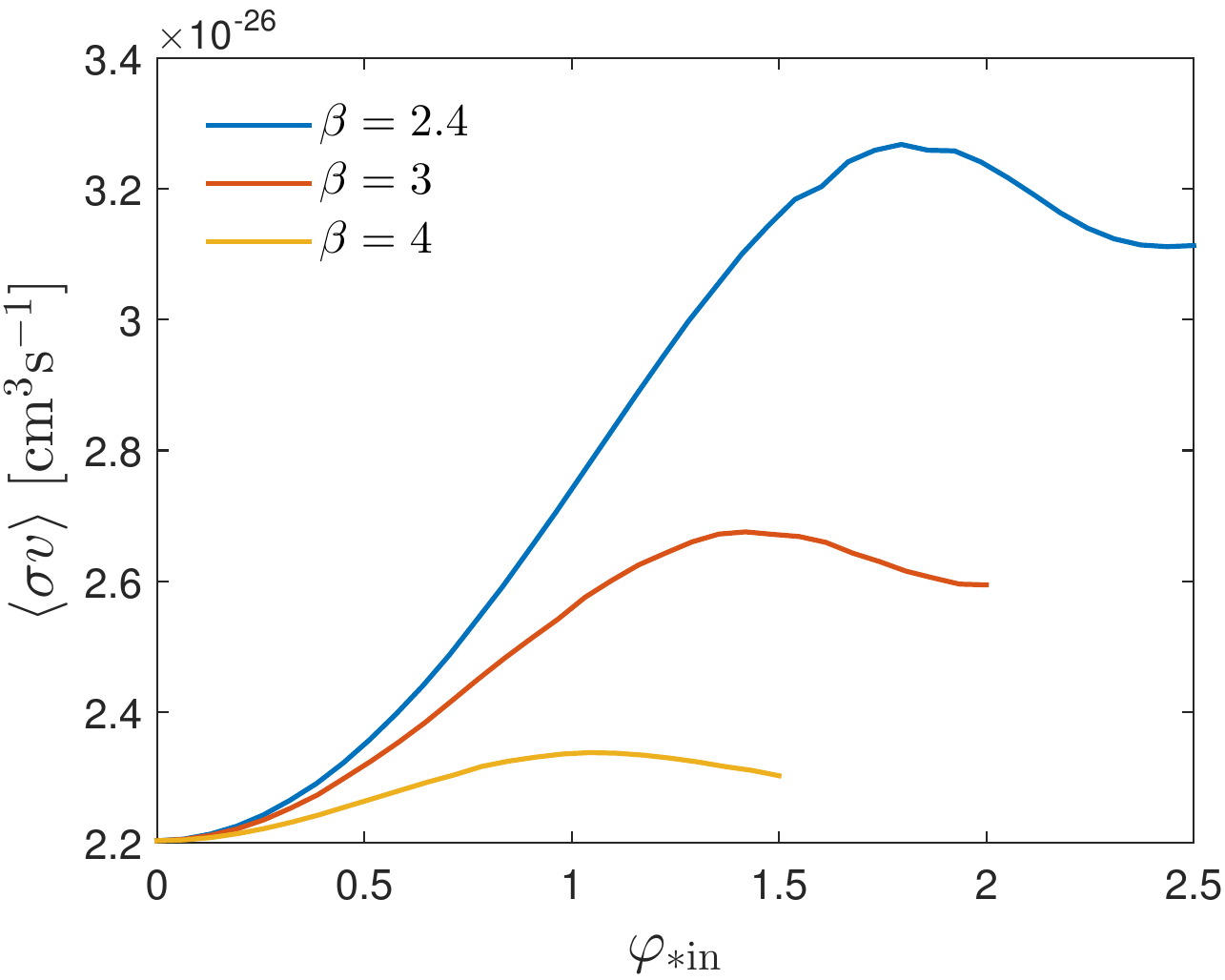}
\hfill
\includegraphics[scale=0.53,trim=0 0 0 0,clip=true]{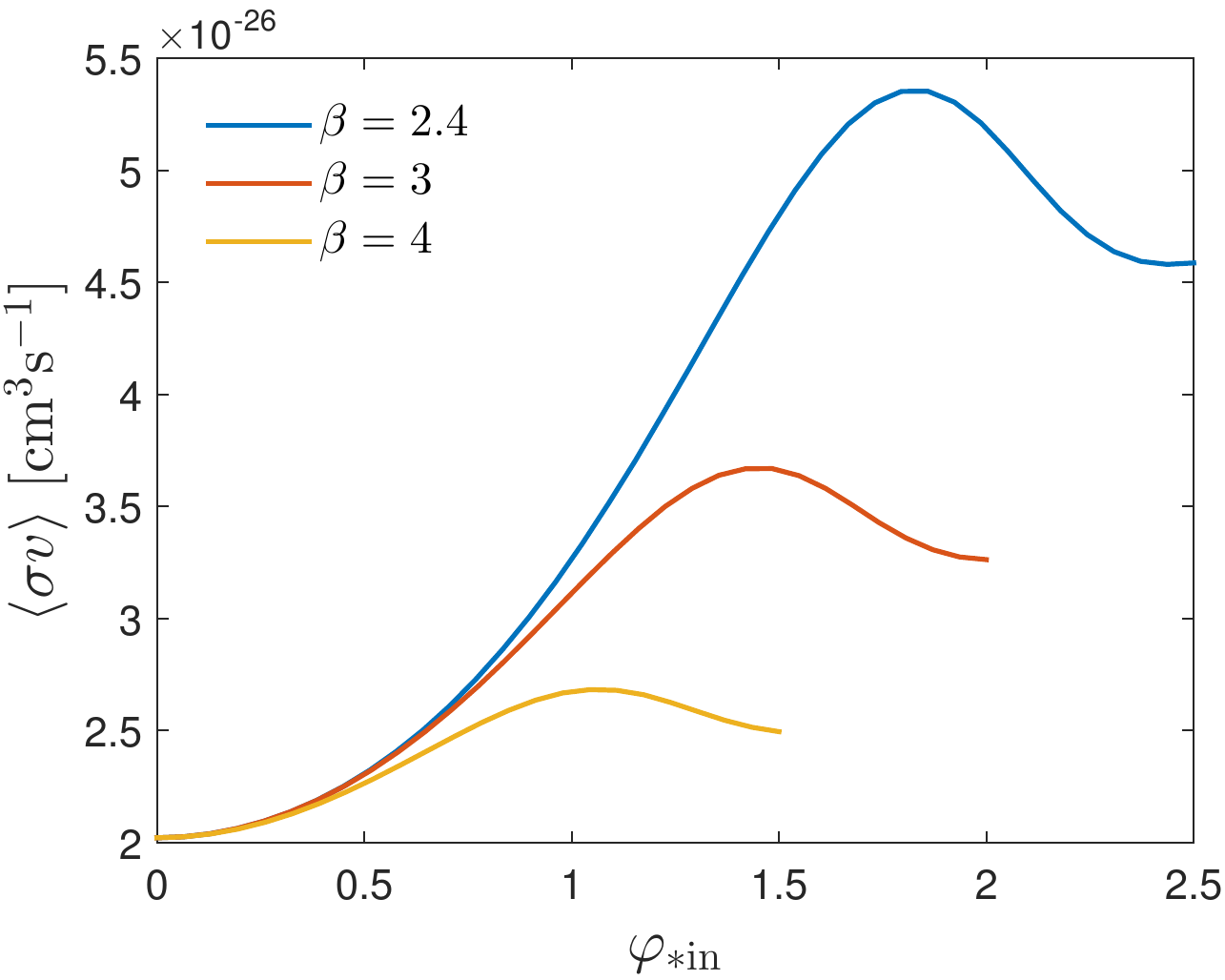}
\caption[Required annihilation cross section in scalar-tensor gravity model ($s-$wave)]{\label{fig:sigma_swave_st} Required $s-$wave annihilation cross section $\langle\sigma v\rangle$ for symmetric dark matter for  $m_\chi = 10$ GeV (left panel) and $m_\chi = 100$ GeV (right panel) as a function of $\varphi_{*{\mathrm{in}}}$ for different values of $\beta$.}
\end{figure}

\begin{figure}[tbp]
\centering
\includegraphics[scale=0.53,trim=0 0 0 0,clip=true]{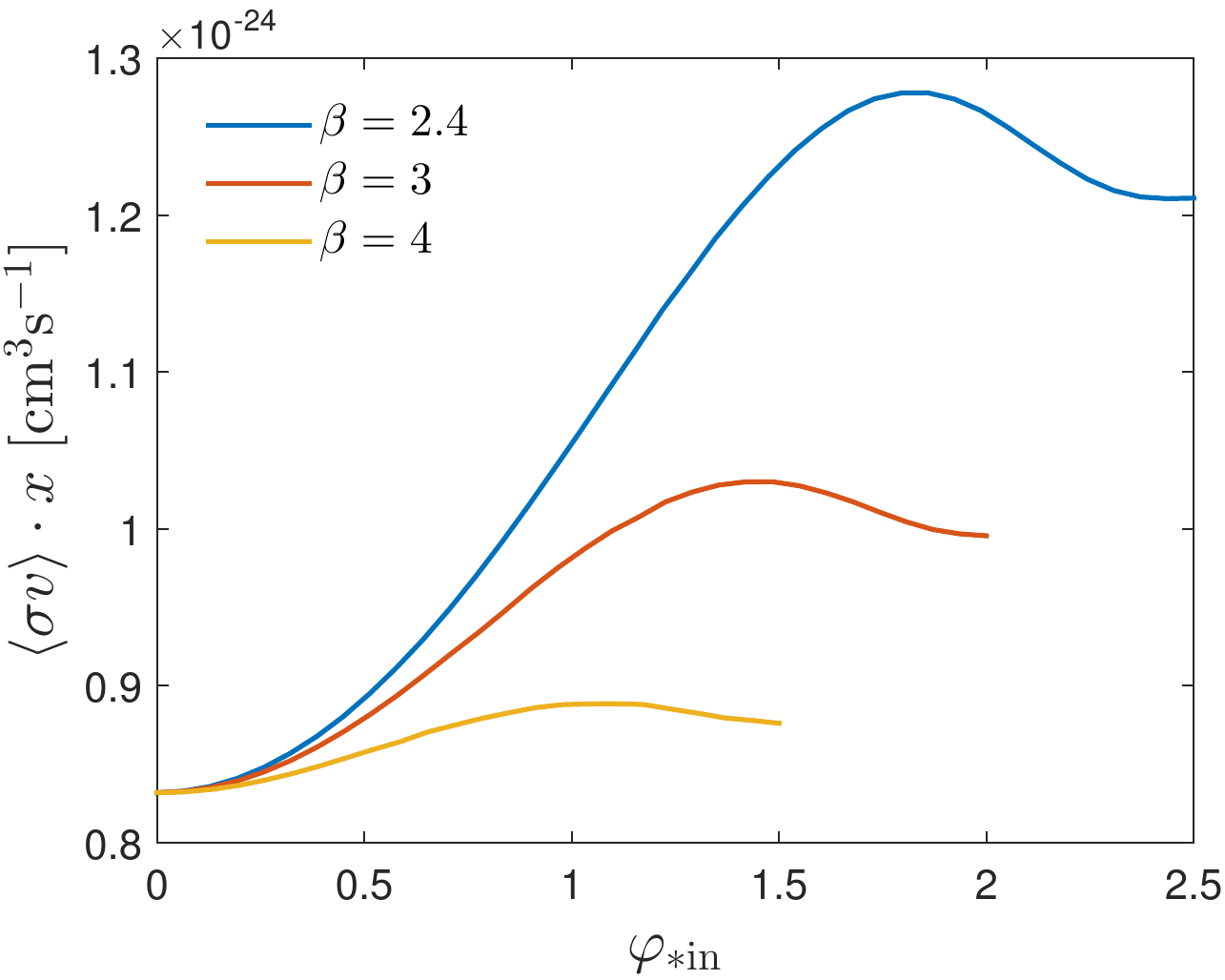}
\hfill
\includegraphics[scale=0.53,trim=0 0 0 0,clip=true]{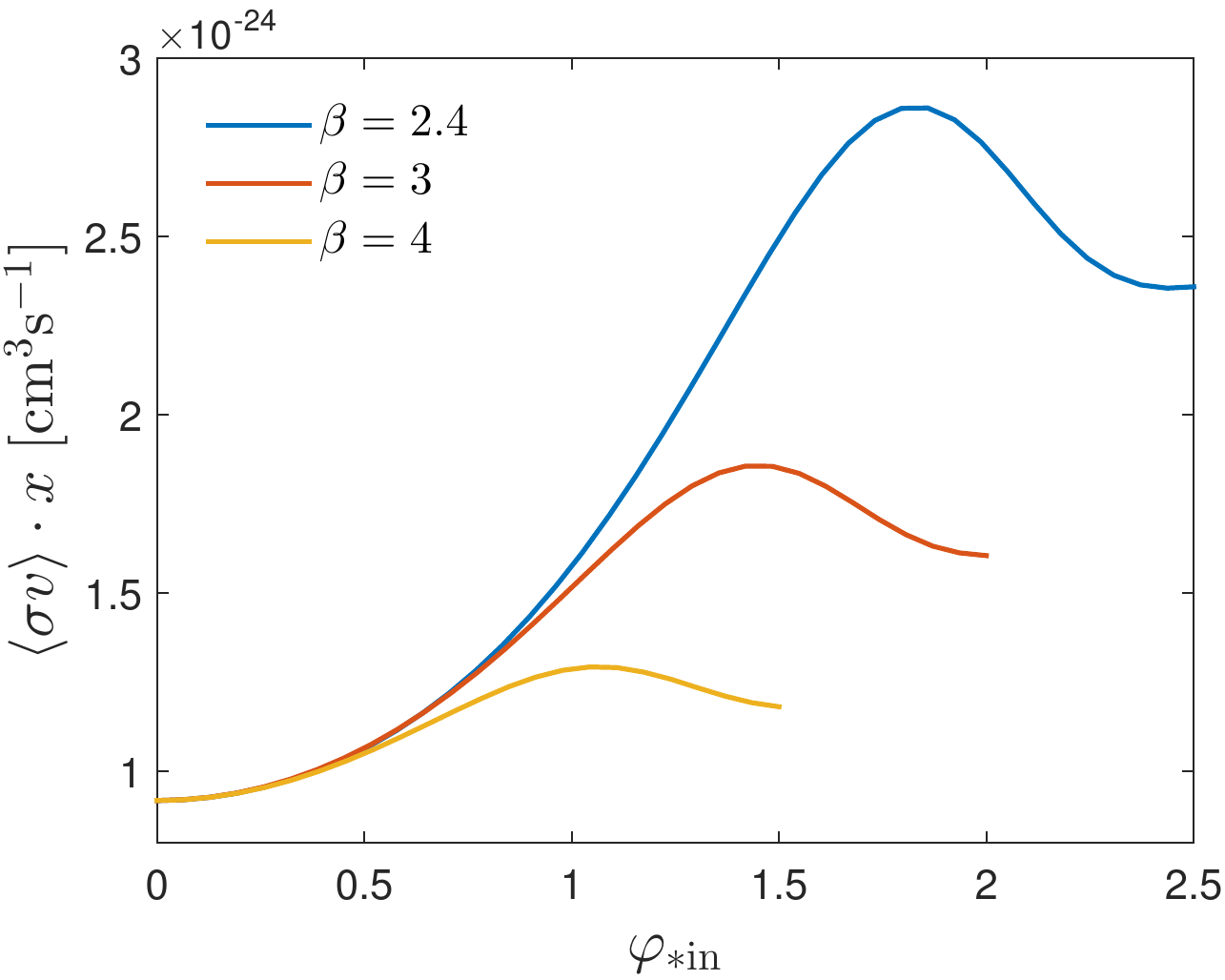}
\caption[Required annihilation cross section in scalar-tensor gravity model ($p-$wave)]{\label{fig:sigma_pwave_st} Same as figure~\ref{fig:sigma_swave_st} but for $p-$wave annihilation.}
\end{figure}

Using the observed dark matter density~\eqref{eq:dm_abun} we can invert our calculation to determine the required annihilation cross section. The results are plotted in figures~\ref{fig:sigma_swave_st} and~\ref{fig:sigma_pwave_st} for the $s-$ and $p-$wave annihilation cases respectively and show similar behaviour to the relic density curves given in figure~\ref{fig:coc_universal_omega}.

As in~\cite{Meehan:2014zsa,Meehan:2014b} we could compare our results with the Fermi-LAT data~\cite{Fermi} which places an upper limit on the dark matter annihilation cross section. However, since, in this case, the enhancement factors are so small, the Fermi-LAT bounds would not be able to significantly constrain the allowed model parameters.


\section{Asymmetric dark matter}
\label{sec:st_adm}

For completeness we also determine the relic abundance of asymmetric dark matter particles in scalar-tensor gravity. Since the symmetric annihilation cross section is only enhanced slightly, we expect a similarly small enhancement for the asymmetric annihilation cross section. 

Asymmetric dark matter models treat the dark matter particle $\chi$ and antiparticle $\bar{\chi}$ as distinct and with unequal number densities, similar to the asymmetry that exists in the baryonic sector~\cite{Petraki:2013wwa,Kumar:2013vba,Zurek201491}. These models typically assume~\cite{Graesser:2011wi} either a primordial asymmetry in one sector that is transferred to the other sector, or that both asymmetries are generated by the same physical process such as the decay of a heavy particle. Relating the asymmetry in the dark matter sector to that in the baryonic sector also explains the proximity of the dark and baryonic densities, $\Omega_{\mathrm{DM}}/\Omega_b \sim 5$, suggesting the dark matter mass is in the range $m_\chi \sim 5 - 15$ GeV~\cite{PhysRevD.79.115016}.

For distinct particle $\chi$ and antiparticle $\bar{\chi}$, the Boltzmann equation~\eqref{eq:boltsym} is generalized to the coupled system
\begin{subequations}
\begin{eqnarray}
\frac{dn_\chi}{dt} &=- 3Hn_\chi -\langle\sigma v\rangle\left(n_\chi n_{\bar{\chi}} - n_\chi^{\mathrm{eq}}n_{\bar{\chi}}^{\mathrm{eq}}\right),\label{eq:nchi} \\
\frac{dn_{\bar{\chi}}}{dt} &= - 3Hn_{\bar{\chi}} -\langle\sigma v\rangle\left(n_\chi n_{\bar{\chi}} - n_\chi^{\mathrm{eq}}n_{\bar{\chi}}^{\mathrm{eq}}\right)\label{eq:nchibar},
\end{eqnarray}
\label{eq:nchiasym}
\end{subequations}
where $n_\chi^{\mathrm{eq}}$ and $n_{\bar{\chi}}^{\mathrm{eq}}$ are the equilibrium number densities of the $\chi$ and $\bar{\chi}$ components respectively.
We assume that self annihilations are forbidden, and that only interactions of the type $\chi\bar{\chi}\rightarrow X\bar{X}$ (where the $X$'s are Standard Model particles) can change the dark matter particle number. We can then write
\begin{equation}
Y_\chi - Y_{\bar{\chi}} = C\label{eq:Cdef},
\end{equation}
where $C$ is a strictly positive constant that characterizes the asymmetry between the particles and antiparticles. Here, we are not concerned with the mechanism that generates the asymmetry, only that one has been created well before particle freeze-out.

In terms of the comoving densities $Y_{\chi}$ and $Y_{\bar{\chi}}$, the Boltzmann equations~\eqref{eq:nchiasym} become
\begin{align}
\frac{dY_{\chi}}{dx}&=-\frac{s\langle\sigma v\rangle}{xH}\zeta(x)\left(Y_\chi^2 - CY_{\chi} - P\right),\nonumber\\
\frac{dY_{\bar{\chi}}}{dx}&=-\frac{s\langle\sigma v\rangle}{xH}\zeta(x)\left(Y_{\bar{\chi}}^2 + CY_{\bar{\chi}} - P\right),\label{eq:dYasym}
\end{align}
where, since the dark matter particles and antiparticles are non-relativistic at decoupling,
\begin{equation}
P \equiv Y_\chi^{\mathrm{eq}}Y_{\bar{\chi}}^{\mathrm{eq}} = \left(\frac{0.145\,g_\chi}{g_{*}}\right)^2x^3e^{-2x}.\label{eq:Pdef}
\end{equation}
Solving the system~\eqref{eq:dYasym} in the asymptotic limit, the total dark matter density, $\Omega_{\mathrm{DM}}h^2$, is the sum of the $\chi$ and $\bar{\chi}$ components,
\begin{equation}
\Omega_{{\mathrm{DM}}}h^2 = 2.75\times 10^8\,m_\chi\left(Y_\chi^\infty + Y_{\bar{\chi}}^\infty\right).\label{eq:dmtot}
\end{equation}

\begin{figure}[tbp]
\centering
\includegraphics[scale=0.53,trim=0 0 0 0,clip=true]{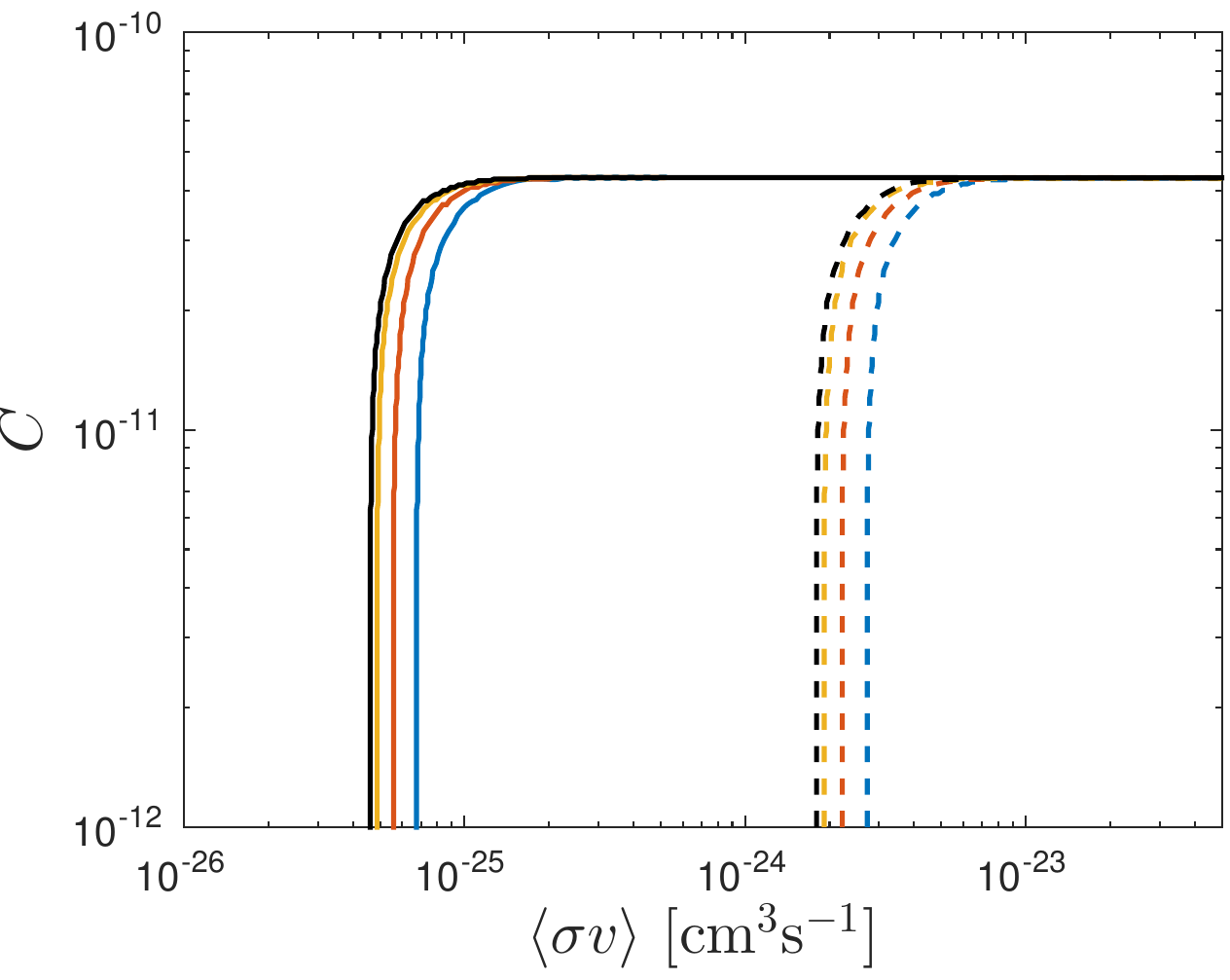}
\hfill
\includegraphics[scale=0.53,trim=0 0 0 0,clip=true]{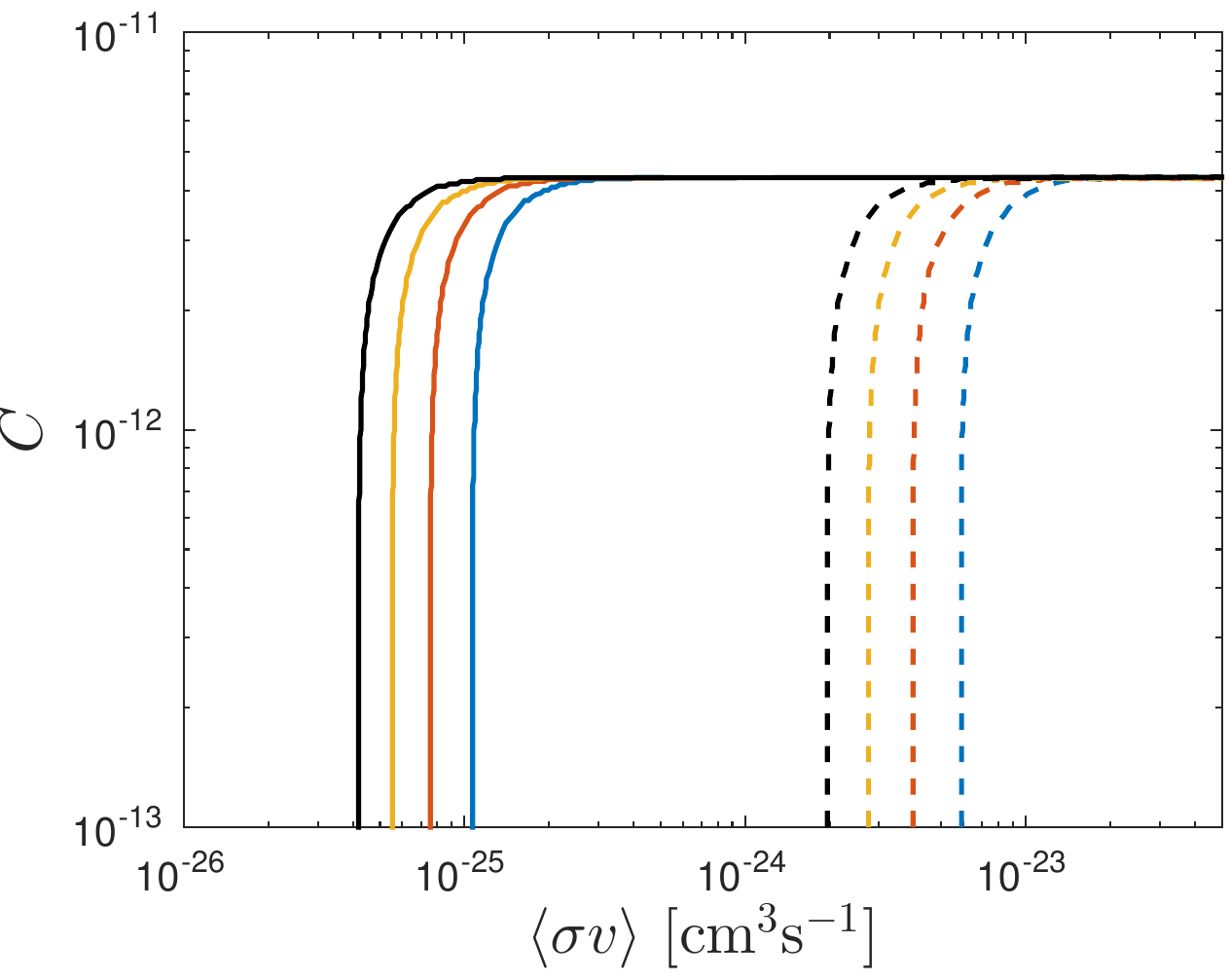}
\caption[Iso-abundance contours in the $(\langle\sigma v\rangle,C$) plane in the scalar-tensor gravity scenario]{\label{fig:asymmetric_sigma_st} Iso-abundance contours in the $(\langle\sigma v\rangle,C)$ plane for asymmetric dark matter corresponding to the observed dark matter abundance $\Omega_{\mathrm{DM}}h^2 = 0.1188$ for $m_\chi =  10$ GeV (left panel) and $m_\chi = 100$ GeV (right panel). The blue, red and yellow curves correspond to the input parameters $(\beta,\varphi_{*\mathrm{in}}) = (2.4,2.0),(3.0,2.0)$ and $(4.0,1.5)$ respectively and the black curves correspond to the standard cosmology result. We have calculated the results for both $s-$ (solid curves) and $p-$ (dashed curves) wave annihilation.}
\end{figure}

In figure~\ref{fig:asymmetric_sigma_st} we plot the iso-abundance contours in the $(\langle\sigma v\rangle,C)$ plane corresponding to the observed dark matter density $\Omega_{\mathrm{DM}}h^2 = 0.1188$. The results have been calculated for both $m_\chi = 10$ GeV (left panel) and $m_\chi = 100$ GeV (right panel) for both the $s-$ (solid curves) and $p-$ (dashed curves) wave cases.

Because the vertical asymptote is only slightly shifted from the standard cosmology result (black curve), we do not expect the detection signal from annihilation of asymmetric dark matter to differ appreciably in the scalar-tensor gravity scenario. However, because the curves are still shifted, it is possible to have a stronger asymmetric detection signal in the scalar-tensor scenario compared with the symmetric signal in the standard scenario (see e.g.~\cite{Gelmini:2013awa,Meehan:2014zsa,Meehan:2014b}) .

The relic abundance of asymmetric dark matter in scalar-tensor gravity has very recently been studied by Gelmini~\textit{et al}~\cite{Gelmini:2013awa} and Wang~\textit{et al}~\cite{Wang:2015gua}. However these studies are based upon the enhancement factor~\eqref{eq:xi_catena} and also neglect the temperature dependence of $g_*(T)$. Unsurprisingly, our results are much less dramatic than those given by Gelmini~\textit{et al}~\cite{Gelmini:2013awa} and Wang~\textit{et al}~\cite{Wang:2015gua} who found that the required annihilation cross section can be increased by several orders of magnitude.



\section{Non-universal scalar-tensor theories}
\label{sec:st_nonuniversal}


\subsection{Visible and Dark Jordan frames}

The formalism and results derived thus far only apply for the special class of scalar-tensor theories known as \textit{universal} scalar-tensor theories in which each of the different matter fields experiences the same coupling $A(\varphi_*)$. Here we consider a more general arrangement where the coupling to each matter sector is distinct. Specifically, we will consider the case studied by Coc~\textit{et al} 2009~\cite{PhysRevD.79.103512} where the coupling between the scalar field and the \textit{visible} (or Standard Model) sector, denoted $A_V(\varphi_*)$, is distinct from that between the scalar field and the \textit{dark} sector, $A_D(\varphi_*)$.\footnote{This set-up is different from that considered by Catena~\textit{et al} in~\cite{1126-6708-2008-10-003} in which they introduced a new \textit{hidden} matter sector and included dark matter within the visible sector.} In this case the Einstein frame action integral~\eqref{eq:ein_act} is generalized to
\begin{align}
S_{\mathrm{tot}} &= \frac{1}{16\pi G_*}\int{d^4x\,\sqrt{-g_*}\left[R_* - 2g_*^{\mu\nu}\partial_\mu\varphi_*\partial_\nu\varphi_* - 4V(\varphi_*)\right]}\nonumber\\
&\qquad + S_V[A_V^2(\varphi_*)g_{\mu\nu}^*\,;\Psi_V] + S_D[A_D^2(\varphi_*)g_{\mu\nu}^*\,;\Psi_D],\label{eq:nonuniversal_action}
\end{align}
where $S_V$ and $S_D$ are the action integrals for the visible, $\Psi_V$, and dark, $\Psi_D$, matter fields respectively. The corresponding generalization of the set of cosmological equations appearing in section~\ref{sec:st_cosmological} to the non-universal case is given in Coc~\textit{et al}~\cite{PhysRevD.79.103512}. We give a brief summary in Appendix~\ref{A}. Here we discuss the implications for the derivation of the Boltzmann equation.

For the general case $A_V\neq A_D$ it is not possible to transform to a conformal frame in which both the visible and dark sectors couple directly to the same metric, that is, there is no unique definition of the Jordan frame. Therefore we must distinguish between two separate Jordan frames: the Visible Jordan Frame (VJF) defined by the metric
\begin{equation}
g_{\mu\nu}^V = A_V^2(\varphi_*)g_{\mu\nu}^*
\end{equation}
and the Dark Jordan Frame (DJF) defined by
\begin{equation}
g_{\mu\nu}^D = A_D^2(\varphi_*)g_{\mu\nu}^* = B(\varphi_*)g_{\mu\nu}^V.\label{eq:dark2vis}
\end{equation}
In the VJF and DJF respectively, the visible and dark sector matter fields do not experience the scalar coupling and particle properties (e.g. mass, energy, cross sections) and their interactions take their standard form.

\subsection{Boltzmann equation}

Before attempting to calculate the dark matter relic abundance, we should first pause to contemplate whether, in the non-universal coupling case, the assumptions made in deriving the Boltzmann rate equation~\eqref{eq:boltsym} (see e.g.~\cite{KandT}) remain valid.

Starting with the fundamental form of the Boltzmann equation,
\begin{equation}
\hat{L}[f(x^\mu,p^\mu)] = \hat{C}[f(x^\mu,p^\mu)]\label{eq:bolt_st_nu}
\end{equation}
where $f(x^\mu,p^\mu)$ is the dark matter distribution function, $\hat{L}$ is the relativistic Liouville operator and $\hat{C}$ is the collision operator, the first question to address is in which frame should we formulate the problem? Given that the subject of the calculation, the dark matter particles, couple directly to the metric $g_{\mu\nu}^D$, the obvious answer is the DJF. Then, proceeding as in~\cite{KandT} we can determine the zeroth moment of~\eqref{eq:bolt_st_nu} by integrating both sides of the equation over momentum space. 

To evaluate the integral of the Liouville operator, we note that the conformal transformation~\eqref{eq:dark2vis} conserves both the isotropy and homogeneity of the VJF metric $g_{\mu\nu}^V$ so that the dark matter distribution function, which is defined in the DJF, reduces to $f(x^\mu,p^\mu) = f(E,t)$.\footnote{Observations of the isotropy and homogeneity of the universe are made with Standard Model particles and therefore apply to the VJF.}$^,$\footnote{In general, a conformal transformation is a \textit{local} rescaling of the metric,
\begin{equation}
\tilde{g}_{\mu\nu} = \Omega^2(x)g_{\mu\nu},
\end{equation}
that only preserves the isotropy of spacetime and not homogeneity. But, if the conformal factor is a function of time only, i.e. $\Omega(x^\mu) \equiv \Omega(t)$, as in~\eqref{eq:dark2vis}, then homogeneity is conserved also.} Therefore, we recover the standard expression
\begin{equation}
\frac{g}{(2\pi)^3}\int{\hat{L}[f(E,t)]\frac{d\vec{p}}{E}} = \frac{dn_D}{dt_D} + 3H_Dn_D
\end{equation}
where $n_D \equiv n_D(t_D)$ and $t_D$ are the dark matter number density and time as defined in the DJF.

However, the evaluation of the collision term in~\eqref{eq:bolt_st_nu} is not so straightforward. The change in the dark matter particle number is governed by reactions of the type 
\begin{equation}
\chi\bar{\chi} \leftrightarrow X\bar{X}\label{eq:chi_reaction}
\end{equation}
where the $\chi$'s are the dark matter particles and the $X$'s are particles belonging to the visible sector. In the non-universal case the particles on either the left or right hand side of the reaction~\eqref{eq:chi_reaction} will either be susceptible or immune to the scalar interaction depending on the choice of conformal frame of reference; there is no conformal frame in which the scalar coupling vanishes for both sides of the reaction. Therefore, if we decide to formulate the Boltzmann equation in the DJF we must account for the varying $\varphi_*$-dependent masses, energies, cross sections, etc. of the visible particles. Of course, transforming to the VJF does not alleviate the problem because then dark matter particles would be $\varphi_*$-dependent.

A proper analysis of the problem would require a thorough reexamination of the derivation of the Boltzmann equation and in particular the evaluation of the collision integral. Since this is beyond the scope of the present study we leave it as a suggestion for future work.





\section{Conclusions}
\label{sec:st_summary}

The inherent attraction mechanism exhibited by many scalar-tensor gravity models towards General Relativity allows for deviations from the standard cosmological scenario in the early universe that may not show up in present observational data. In fact, it has been conjectured that relic abundance calculations may be one of the few probes capable of discriminating the predictions of scalar-tensor scenarios from standard General Relativity~\cite{Gelmini:2009yh}. To find out, we determined the evolution of the coupled scalar field from first principles, allowing us to calculate the modified expansion rate in the scalar-tensor gravity scenario, which we then used to calculate the dark matter relic abundance. 

As a specific example we considered the prototypical quadratic coupling $A(\varphi_*) = e^{\frac{1}{2}\beta\varphi_*^2}$ and found that the maximum enhancement for a $m_\chi = 100$ GeV WIMP was $\Omega_{\mathrm{DM}}^{\mathrm{ST}}/\Omega_{\mathrm{DM}}^{\mathrm{GR}} \sim 3$. Although this ratio would increase with increasing WIMP mass, the level of enhancement is still far below that found in previous relic abundance investigations in scalar-tensor cosmology~\cite{PhysRevD.70.063519,PhysRevD.81.123522}.

Although the expansion rate at the time of dark matter decoupling in the scalar-tensor scenario can be up to several orders of magnitude larger than the expansion rate in the standard scenario (see figure~\ref{fig:dilaton_xi_fo}), these points were excluded by BBN constraints~\cite{PhysRevD.73.083525}; had these points been acceptable, we would have found much larger relic density enhancement factors, possibly in line with those reported in~\cite{PhysRevD.70.063519} and~\cite{PhysRevD.81.123522}. Given that detailed BBN constraints for the couplings $A(\varphi_*) = 1 + Be^{-\beta\varphi_*}$ and $A(\varphi_*) = 1 + b\varphi_*^2$, considered in~\cite{PhysRevD.70.063519} and~\cite{PhysRevD.81.123522} respectively, were not applied in their investigations because a study equivalent to~\cite{PhysRevD.73.083525} was not available, we suggest that the allowed expansion rate in those models may be much smaller than previously reported, and that the level of enhancement of the dark matter relic abundance might actually be much closer to the levels found here.


To complete our study we also investigated the relic abundance of asymmetric dark matter species in scalar-tensor gravity models and found that the required annihilation cross section was enhanced only slightly, in contrast to the several order of magnitude estimates given in Gelmini~\textit{et al}~\cite{Gelmini:2013awa} and Wang~\textit{et al}~\cite{Wang:2015gua} who both used the parameterization~\eqref{eq:xi_catena} given in Catena~\textit{et al}~\cite{PhysRevD.70.063519}.

Finally, since the attraction mechanism towards GR is typically initiated prior to dark matter freeze-out so that the effect on the dark matter relic abundance is only modest, the scalar-tensor scenario with a quadratic coupling to matter does not provide a significantly different picture from the standard cosmological scenario.
Hence, we conclude that unless reasonably precise details about the nature of the dark matter particle and its interactions are known, relic abundance calculations are most likely unable to discriminate scalar-tensor gravity from General Relativity.

\appendix
\section{Non-universal scalar-tensor theories: Cosmological equations}
\label{A}
The cosmological equations for a non-universal scalar-tensor theory for which the coupling between the visible ($V$) and dark ($D$) sectors is distinct, i.e. $A_V\neq A_D$, has been given previously in, e.g. Coc~\textit{et al}~\cite{PhysRevD.79.103512}. We briefly summarize the main results here. 

Following the prescription given in section~\ref{sec:conformal_frames}, we can apply the conformal transformation
\begin{equation}
g^V_{\mu\nu} = A^2_V(\varphi_*)g_{\mu\nu}^*
\end{equation}
to remove the coupling from the visible sector matter action so that the fields $\Psi_V$ couple to the metric $g_{\mu\nu}^V$ directly. The action integral~\eqref{eq:nonuniversal_action} then becomes
\begin{align}
S_{\mathrm{tot}} &= \frac{1}{16\pi G_*}\int{d^4x\,\sqrt{-g_V}\left[F(\varphi)R_V - g_V^{\mu\nu}Z(\varphi)\partial_\mu\varphi\partial_\nu\varphi - 2U(\varphi)\right]}\nonumber\\
&\qquad  + S_V[g^V_{\mu\nu}\,;\Psi_V] + S_D[B^{2}(\varphi_*)g^V_{\mu\nu}\,;\Psi_D];\label{eq:vis_jord_act}
\end{align}
where the connection between $\varphi$ and $\varphi_*$, and the definitions of the functions $F(\varphi),\,Z(\varphi)$ and $U(\varphi)$, are given in~\eqref{eq:field_redef} with $A(\varphi_*)$ replaced by $A_V(\varphi_*)$. Also, we have defined
\begin{equation}
B(\varphi_*) = \frac{A_D(\varphi_*)}{A_V(\varphi_*)}.
\end{equation}
Although the visible sector action $S_V[g_{\mu\nu}^V\,;\Psi_V]$ is independent of the scalar field in this new frame, the term $S_D[B^{2}(\varphi_*)g_{\mu\nu}^V\,;\Psi_D]$ still contains a scalar coupling. We must therefore distinguish between two distinct Jordan frames: the Visible Jordan Frame (VJF) defined by the metric $g_{\mu\nu}^V$, and the Dark Jordan Frame (DJF) defined by
\begin{equation}
g_{\mu\nu}^D = A_D^2(\varphi_*)g_{\mu\nu}^* = B(\varphi_*)g_{\mu\nu}^V.
\end{equation}
Similar to~\eqref{eq:alpha}, we can define the parameters that characterize the deviation from General Relativity as
\begin{equation}
\alpha_j(\varphi_*) = \frac{d\ln A_j(\varphi_*)}{d\varphi_*},
\end{equation}
where $j = V,D$ labels the different matter sectors.

In the non-universal case the cosmological equations~\eqref{eq:ein_hub} and~\eqref{eq:ein_hub2} for the evolution of the Einstein frame scale factor $a_*$ remain unaltered. However, the matter fields of each sector obey separate continuity equations
\begin{equation}
\frac{d\rho_{*j}}{dt_*} + 3H_*\left(\rho_{*j} + p_{*j}\right) = \alpha_j(\varphi_*)\left(\rho_{*j} - 3p_{*j}\right)\dot{\varphi}_*.\label{eq:nonuniversal_continuity}
\end{equation}
This implies that the visible (dark) sector matter fields are no longer conserved in the DJF (VJF) so that, for example, in the VJF we have 
\begin{align}
\rho_{Vi} &= \rho_{Vi}^0\left(\frac{a_V}{a_{V0}}\right)^{-3(1 + w_i)},\nonumber\\
\rho_{D} &= \rho_{D}^0\left[\frac{B(\varphi_*)}{B(\varphi_{*0})}\right]^{4 - 3(1 + w_D)}\left(\frac{a_V}{a_{V0}}\right)^{3(1 + w_D)},
\end{align}
where $i$ labels the different visible fluid components and $\rho_D$ is the dark matter energy density as measured in the VJF. 

Moreover, the equation of motion~\eqref{eq:ein_phimotion} for the scalar field is generalized to
\begin{equation}
\ddot{\varphi}_* + 3H_*\dot{\varphi}_* + \frac{\partial V}{\partial \varphi_*} = -4\pi G_* \sum_{j = V,D} \alpha_j\left(\rho_{*j} - 3p_{*j}\right).
\end{equation}
Finally, the connection between the Einstein frame expansion rate and the VJF and DJF expansion rates is given by
\begin{align}
H_j = A^{-1}_j(\varphi_*)\left[H_* + \alpha_j(\varphi_*)\dot{\varphi}_*\right].
\end{align}
Coc~\textit{et al} 2009~\cite{PhysRevD.79.103512} have studied the BBN constraints on models with quadratic couplings 
\begin{equation}
A_j(\varphi_*) = e^{\frac{1}{2}\beta_j\varphi_*^2} \qquad (j = V,D).
\end{equation}
In particular, they studied regions in the $(\beta_V,\beta_D)$ parameter space for which there is late time attraction towards GR and then derived additional constraints on the parameter space from BBN and precision gravitational tests.

\end{document}